\documentclass[pre]{revtex4}
\usepackage{graphicx}
\usepackage{amsmath}
\usepackage{bbold}
\usepackage{amssymb}
\usepackage{color}
\usepackage[normalem]{ulem}
\usepackage{epsfig}
\usepackage{amssymb,amsfonts,amsmath,color}
\usepackage{array}
\usepackage{makecell}

\begin{document}

\title{Taming the diffusion approximation through a controlling-factor WKB method}

\author{Jayant Pande and Nadav M.\ Shnerb}

\affiliation{Department of Physics, Bar-Ilan University, Ramat-Gan IL52900, Israel}

\begin{abstract}
\noindent The diffusion approximation (DA) is widely used in the analysis of stochastic population dynamics, from population genetics to ecology and evolution. The DA is an uncontrolled approximation that assumes the smoothness of the calculated quantity over the relevant state space and fails when this property is not satisfied. This failure becomes severe in situations where the direction of selection switches sign. Here we employ the WKB (large-deviations) method, which requires only the logarithm of a given quantity to be smooth over its state space. Combining the WKB scheme with asymptotic matching techniques, we show how to derive the diffusion approximation in a controlled manner and how to produce better approximations, applicable for much wider regimes of parameters. We also introduce a scalable (independent of population size) WKB-based numerical technique. The method is applied to a central problem in population genetics and evolution, finding the chance of ultimate fixation in a zero-sum, two-types competition.
\end{abstract}

\maketitle

\section{Introduction}

Populations  -- collections of individuals which involve some kind of a birth and death process -- are a fundamental object of study throughout the biological sciences. The number of individuals in a population is by definition an integer, and the birth-death process is inherently stochastic. As a result, the study of population dynamics -- in genetics, ecology and evolution -- requires one to examine stochastic processes over the set of integers~\cite{crow1970introduction,ewens2012mathematical,lande2003stochastic}.

These processes are characterized by different rates: the birth rate, for example, reflects the chance (per unit of time) of a given individual to produce an offspring. When the rates are fixed in time the stochastic process is stationary and the noise is binomial: for a population of $n$ individuals of a given species the standard deviation of the total number of offspring per unit time is proportional to $\sqrt{n}$. The rates themselves vary in time when the reproductive success of entire populations fluctuates coherently as a result of macro-environmental changes. If these variations are stochastic, which is a reasonable assumption given the complexity of all biological environments, the amplitude of abundance variations scales with $n$. The ${\cal O}(\sqrt{n})$ noise, which is uncorrelated among individuals, is known in the literature as demographic stochasticity, internal noise or genetic drift, while the  ${\cal O}(n)$ noise, originating from macro-variations of the environment, is usually described as external noise, fluctuating selection or environmental stochasticity~\cite{lande2003stochastic,assaf2017wkb,wienand2017evolution}.

Mathematically speaking, such a stochastic process corresponds to a biased random walk over the integers, and the important quantities which are of interest for life scientists, like the probability of fixation or invasion and the persistence time of a system, all have equivalents in the theory of random walks and first passage times~\cite{redner2001guide}. But unlike simple random walks, here (even in a fixed environment) the rate of jumps and the strength of the bias depend on $n$. Therefore, the effective timescale associated with environmental variations (the typical number of birth-death events before an environmental shift, say) reflects both external factors, like the correlation time of the environment, and internal factors, like the abundance of a given population. This phenomenon makes the mathematical analysis rather complicated. Another source of complexity is the need to consider, in a varying environment, both demographic and environmental stochasticity: while the demographic noise is negligible in large populations, it does control the low-$n$ sector and hence its strength dictates the most important processes, extinction and fixation~\cite{lande2003stochastic,danino2016effect}.

The simplest and the most popular tool in the analysis of these systems is the diffusion approximation (DA)~\cite{karlin1981second}. This approximation is based on the expansion of the relevant quantities in a power series and the retention of only the two leading orders, an approach that assumes the smoothness of these quantities over the integers. Since its introduction the diffusion approximation has gained a lot of popularity, as it provides a generic algorithm that reduces problems of this kind to relatively simple, second-order differential equations such as the Fokker-Planck or the  Backward Kolmogorov equation. The central place of the diffusion approximation in the modern theory of population genetics was surveyed by Wakeley~\cite{wakeley2005limits}, who described how, because of its usefulness as a calculation tool, the diffusion approximation has shaped the conceptual framework of the field.

For reasons that are not quite clear, the performance of the diffusion approximation in a fixed environment (when selection is fixed through time and demographic stochasticity is the only noise-producing mechanism) is excellent~\cite{parsons2010some}. On the other hand, in the presence of environmental stochasticity and fluctuating selection -- for the importance of which there is ever-increasing evidence in the literature~\cite{bergland2014genomic,bell2010fluctuating,messer2016can,caceres1997temporal,hoekstra2001strength,leigh2007neutral,hekstra2012contingency,kalyuzhny2014niche,kalyuzhny2014temporal,chisholm2014temporal} -- the performance of the DA, as we shall see below, is much poorer. This poses a severe challenge to theory: most of the old~\cite{takahata1975effect,takahata1979genetic} and new~\cite{huerta2008population,hidalgo2017species,danino2018stability,danino2018fixation,meyer2018noise,danino2018theory,meyer2020evolutionary} studies of the effects of fluctuating selection or environmental stochasticity are based on the diffusion approximation, but the parameter range in which this generic technique is applicable turns out to be rather narrow.

The existing alternatives to the DA for systems with fluctuating selection have their own problems. Numerical surveys, as in~\cite{ashcroft2014fixation}, are case-specific and are limited to relatively small systems. The applicability of Haldane's branching process approximation~\cite{uecker2011fixation,engen2009reproductive,marrec2020resist} is restricted to small densities (it cannot take into account density-dependent, nonlinear effects) and its use in a system with a varying environment is technically complicated. Other works use heuristic arguments that rely on an interpolation between the diffusive and the nearly-fixed regimes~\cite{mustonen2008molecular,cvijovic2015fate,wienand2017evolution,wienand2018eco}. While these may be effective in particular scenarios, a general method to treat diverse systems under (weak as well as strong) fluctuating selection -- one that may take the place of the DA -- is still lacking.

Here we present just such a generic approach, by combining the large-deviations (WKB) theory with asymptotic matching techniques. Our approach may be adopted for a wide variety of problems in population genetics, ecology and evolution where the sign and the amplitude of the selection vary stochastically in time.

Large-deviations theory has been applied in the past to continuum (no demographic noise) systems under large-amplitude external noise~\cite{kubo1973fluctuation,gang1987stationary,dykman1994large} and to systems with only demographic stochasticity (intrinsic noise)~\cite{kessler2007extinction}. The combined effect of both types of stochasticity has been studied when the environmental fluctuations reflect an underlying Ornstein-Uhlenbeck process~\cite{kamenev2008colored,levine2013impact,assaf2017wkb} or are modeled as an auxiliary species~\cite{assaf2008noise,roberts2015dynamics}.  Under these conditions the problem becomes two-dimensional and must be solved numerically.

We would like to break free of the limitations in existing approaches by treating generic temporal environmental fluctuations as a one-dimensional problem and combining the WKB method with the asymptotic matching technique that was employed with great success for the diffusion approximation~\cite{danino2018stability,danino2018fixation,meyer2018noise,yahalom2019comprehensive,yahalom2019phase}.
We consider the common case in which the fixation time is much larger than the correlation time of the environment (this is the ``annealed" case of~\cite{mustonen2008molecular}; the chance of fixation otherwise depends strongly on the initial state of the environment). This allows us to average over different states of the environment and to map the problem onto a biased, one-dimensional walk, where step sizes and directions reflect effective selection and effective (both demographic-induced and environmental) stochasticity. The WKB method then yields a simple, single-parameter transcendental equation for which we present efficient approximate solutions. These solutions are employed separately in the inner regime (close to extinction), in the outer regime (close to fixation), and in the middle regime (where the demographic noise is negligible with respect to the environmental variations), and are matched to each other in the regions of overlap.

The range of applicability of this generic approach turns out to be much wider than that of the DA, and it converges to the DA when a given parameter is small, so its usage makes the DA technique a controlled approximation. Moreover, our approach allows further improvements, such as by adding more regimes to the asymptotic matching procedure or by modifying the transcendental equation to reflect better the details of the underlying process. We also show how to employ our method as a scalable ($N$-independent, where $N$ is the total population of the entire community counting all the species) numerical technique that works even when the simple asymptotic analysis fails.

To clarify the discussion we stick here to the calculation of the simplest and the most important quantity in the theory of population genetics and evolution, namely the chance of ultimate fixation. The environmental stochasticity is modeled by erratic fluctuations between two states (dichotomous noise), as explained below. These assumptions impose no restrictions on the utilization of the technique presented here, which, \emph{mutatis mutandis}, may be used to calculate other quantities (like the time to fixation, the time to absorption and the quasi-stable distribution function) under different realizations of stochasticity.

This paper is organized as follows. We begin, in the next section, with the case of fixed selection. The DA works well in this case, so the goal of the discussion here is didactic: it allows us to present two critical elements of our method, the WKB technique and the two-destination approximation, and to demonstrate our ability to derive \emph{the same results} using the DA and using the new technique. The two-destination approximation, that we present here for the first time, yields naturally what we term the fundamental transcendental equation, and its approximate solutions in different parameter regimes (sectors) are presented in Section~\ref{fundamental}.

Section~\ref{model1} defines the stochastic dynamics of systems with fluctuating selection, and explains how to solve numerically for the chance of ultimate fixation using matrix inversion.
In section~\ref{s7} we employ the solutions presented in Section~\ref{fundamental} and establish a scalable numerical scheme where (unlike the direct numerics which involve the inversion of an $N \times N$ matrix) the numerical effort is almost $N$-independent. To obtain an analytical solution we have to employ a large-$N$ asymptotic matching technique. This method depends on the existence of a middle regime where the demographic stochasticity is negligible. We clarify the conditions (on the magnitude of $N$) for such a middle regime to exist and further discuss the technique in Section~\ref{s3}. An analytical solution is first presented in Section~\ref{apF}. We show how to derive the DA result (as obtained in~\cite{danino2018fixation,meyer2018noise}) as a limiting case of this more general solution, demonstrate the superiority of the WKB solution with respect to the DA result, and use it to clarify the important distinction between weak and strong selection under a fluctuating environment. Subsequently, in Section~\ref{lastapp} we extend the results to different parameter regimes. The emerging general picture is reexamined in the discussion section.

\section{Fixed selection} \label{fixed}

Under fixed selection (both positive and negative), the diffusion approximation performs excellently. Our goal in this section is not to improve this approximation but to provide a methodological and technical introduction to the use of the WKB technique. For this purpose we rederive the DA expression, first using the traditional method and then as a limit of the WKB expression.

\subsection{The stochastic process}

We consider a population with $N$ individuals, $n$ mutants and $N-n$ wild-type individuals. If $x=n/N$ is the frequency of the mutants, $1-x$ is the frequency of the wild type. The mutant fitness is $W = e^s$, so $s$, the selection parameter, is the log-fitness. We consider simple haploid non-overlapping generation (Wright-Fisher) dynamics: in each generation  (without loss of generality, assumed to be a year) all the individuals die and their offspring (seed, larvae, etc.) compete for the $N$ empty slots. The probability of the mutant type to capture any given slot is given by
\begin{equation} \label{eq01}
r = \frac{n e^s}{n e^s + (N-n)}.
\end{equation}
Accordingly, the chance of a mutant population of size $n$ to reach, in the next haploid generation, the size $n+m$ ($-n \le m \le N-n$) is
\begin{equation} \label{destinations1}
W_{n \to n+m} = \binom{N}{n+m} r^{n+m}(1-r)^{N-n-m}.
\end{equation}
The mean change in the mutant population size after one year is
\begin{equation}\label{mean}
\mathbb{E}[m] \equiv  \sum_m m W_{n \to n+m} = N(r-x),
\end{equation}
and the variance of this quantity  is
\begin{equation} \label{smom}
\mathbb{E}[m^2] - (\mathbb{E}[m])^2 = N(r-r^2).
\end{equation}

\subsection{The diffusion approximation} \label{DAs}

The chance of ultimate fixation of a mutant species starting from $n$ individuals, $\Pi_n$, satisfies the Backward Kolmogorov Equation,
\begin{equation} \label{BKE}
\Pi_n = \sum_m W_{n \to n+m} \Pi_{n+m}.
\end{equation}
This difference equation admits two independent solutions.  Since $\sum_m W_{n \to n+m} =1$, one solution is always a constant, so the general solution takes the form $\Pi_n = C_1 + C_2 \overline{ \Pi}_n$, where $\overline{ \Pi}_n$ is some non-trivial function of $n$. The boundary conditions,
\begin{equation} \label{bound}
\Pi_0 = 0 \qquad \text{and} \qquad \Pi_N = 1,
\end{equation}
are satisfied by a specific linear combination of  the two solutions, which determines the constants $C_1$ and $C_2$.

The diffusion approximation is based on the assumption that $N$ is large so that $\Pi_n$ is smooth enough over the integers. This allows one to approximate (for $x=n/N$)
\begin{equation} \label{smooth}
\Pi_{n+m} = \Pi(x+m/N) \approx \Pi(x) + \frac{m}{N} \Pi'(x) + \frac{m^2}{2N^2} \Pi''(x) + \frac{m^3}{6N^3} \Pi'''(x) + \rm{higher \ order \  terms},
\end{equation}
where a prime denotes a derivative with respect to $x$. Assuming that $N$ is large and $Ns^2 \ll 1$~\cite{sella2005application}, one finds that (see Appendix~\ref{app:ignoringPi3} for details)
\begin{itemize}
  \item Only the first two terms ($\Pi'$ and $\Pi''$) must be taken into account, and all the higher-order terms (including $\Pi'''$) are negligible.
  \item The difference between the second moment, $\mathbb{E}[m^2]$, and the variance, $\mathrm{Var}[m] \equiv \ \mathbb{E}[m^2] - (\mathbb{E}[m])^2$, is negligible.
\end{itemize}
Accordingly, the diffusion approximation for Eq.~(\ref{BKE}) takes its canonical form,
\begin{equation} \label{canonical}
\frac{\mathrm{Var}[m]}{2N^2} \Pi'' + \frac{\mathbb{E}[m]}{N} \Pi' = 0.
\end{equation}
 Specifically, for the model at hand (which has $\mathbb{E}[m]=s$ and $\mathrm{Var}[m] =1$) one obtains
\begin{equation} \label{eq5}
\Pi''+2sN \Pi'=0, \qquad \text{with} \qquad \Pi(0)=0 \qquad \text{and} \qquad \Pi(1) = 1.
\end{equation}
This has the two linearly independent solutions $C_1$ and $C_2 \exp{-2sN}$, where $C_1$ and $C_2$ are arbitrary constants, so the solution that matches the boundary conditions is the well-known formula~\cite{kimura1962probability,crow1970introduction,ewens2012mathematical},
\begin{equation} \label{sol}
\Pi(x) = \frac{1-e^{-2sNx}}{1-e^{-2sN}}.
\end{equation}

We emphasize that \textbf{the diffusion equation depends on the assumption that $\Pi$ is smooth over the integers}, such that $\Pi_{n+m}$ may be approximated by Eq.~(\ref{smooth}). When $\Pi$ is not smooth enough, the diffusion approximation  fails~\cite{kessler2007extinction}. Note that $N \to \infty$ is not a sufficient condition for smoothness: as may be seen from Eq.~(\ref{sol}), when $s$ is positive (beneficial mutation) the chance of fixation rises from zero at $n = 0$ to almost one when $n=xN=1/s$. The gradient of $\Pi$ may be large even when $N$ diverges.

\subsection{WKB approximation and the two-destination scheme}

Let us now take the WKB approach. As mentioned above, here our aim is not to improve the result in Eq.~(\ref{sol}), but to demonstrate its derivation using the WKB technique and to discuss some details and limitations of WKB as employed here.

WKB is a generic perturbation scheme that allows one to derive a perturbation series that converges (asymptotically) to the correct result. Here we will calculate only the first term (the controlling factor) of this series. At this level the interpretation of our technique is simple: instead of assuming that $\Pi_n$ is smooth over the integers, we assume that its logarithm, $S_n = \ln \Pi_n$, is smooth. It is, of course, possible that even when a function is not smooth enough, its logarithm is.

With the substitution $\Pi_n = e^{S_n}$, Eq.~(\ref{BKE}) takes the form
\begin{align} \label{WKB1}
e^{S_n} = \sum_m W_{n \to n+m} e^{S_{n+m}} \approx  \sum_m  W_{n \to n+m} e^{S(n)+m q(n)}
\end{align}
where $q(n)$ is defined to be $S'(n)$ (for notational convenience) and the approximation above is based on the (assumed) smoothness of $S$.

To continue we would like to get rid of the sum over $m$. We do this by introducing another approximation that has nothing to do with the WKB method itself (apart from simplifying its calculations): \textbf{the two-destination scheme}. This approximation preserves the mean and the variance of the change in a state per timestep, and implicitly rests on the assumption that it is only these first two moments of the change in a state that are important to the dynamics, not the higher moments (see discussion below). To apply this approximation, instead of the random walk described in Eq.~(\ref{destinations1}), we consider an ``effective walk": an asymmetric random walk whose (equally probable) destinations are given by the mean, plus or minus the standard deviation (the square root of the variance). Since our aim for the moment is to reproduce the DA result, Eq.~(\ref{sol}), here we make the same assumptions for the mean and the variance as those used above, namely, that $s$ and $Ns^2$ are small. (Note that when we move beyond the DA in the following sections, we will no longer require these quantities to be very small.)

For $x = n/N$ and $\Delta x \equiv m/N$, we have $\mathbb{E}[\Delta x] \approx sx(1-x)$ and $\sqrt{\mathrm{Var}[\Delta x]} \approx \sqrt{x(1-x)/N}$. The transition probabilities are thus,
\begin{equation} \label{2des}
W_{x \to x+ sx(1-x)+\sqrt{x(1-x)/N}} = 1/2 \qquad \text{and} \qquad W_{x \to x+ sx(1-x)-\sqrt{x(1-x)/N}} = 1/2.
\end{equation}
Plugging (\ref{2des}) into (\ref{WKB1})  one obtains
\begin{equation} \label{WKB2}
e^S =  \frac{e^S}{2} \left(  e^{q(sx(1-x)+\sqrt{x(1-x)/N})} + e^{q(sx(1-x)-\sqrt{x(1-x)/N})}  \right),
\end{equation}
or
\begin{equation} \label{eq10}
 e^{qsx(1-x)} \cosh\left(q\sqrt{\frac{x(1-x)}{N}}\right) = 1.
\end{equation}
Here, and in what follows, we have suppressed the $n$-dependence of $q$ and $S$ (and $\Pi$) for notational convenience, though we will occasionally write them out as $q(n)$, $q(x)$, $S(n)$, $S(x)$, etc.\ (or $q_n$, $S_n$, etc., to highlight the discreteness of $n$) if we want to show the argument explicitly.

As we shall see below, the two-destination approximation yields generically equations that have the general form of  Eq.~(\ref{eq10}). For the moment, let us assume that $q$ is ${\cal O}(1)$ (or, at least, is not a large parameter). In that case, the smallness of $s$ and $1/N$ allows us to expand both the exponent and the cosh functions in a Taylor series to the second order in their arguments, yielding
\begin{align}
\left[1 + q s x (1-x) + \frac{q^2 s^2 x^2 (1-x)^2}{2}\right]\left[1 + \frac{q^2 x (1-x)}{2N}\right] = 1,
\end{align}
which results in the following two solutions for $q$,
\begin{equation} \label{eq11}
q_1 = -2sN \qquad \text{and} \qquad q_2 = 0.
\end{equation}

These two solutions for $q$ yield the two independent solutions for $\Pi$.  Since $S_1 = \int q_1 dx = -2sNx$ and $S_2 = \int q_2 dx = \rm{constant}$, these two independent solutions are $\Pi_1(x) = C_1$ and  $\Pi_2(x) = C_2 \exp(-2sNx)$. The boundary conditions (\ref{bound}) are satisfied if $C_1 = -C_2 = 1/[1-\exp(-2sN)]$, which again yields the solution (\ref{sol}).

\subsection{An outlook} \label{outlook}

As explained above, our goal in this section is not to improve Eq.~(\ref{sol}) but to rederive it using the WKB technique. Let us review the steps involved:
\begin{itemize}
  \item We employ only the controlling factor of the WKB expansion. This approximation has a simple interpretation: it assumes, instead of a smooth $\Pi$, a smooth $S = \ln \Pi$. This approximation may fail: in some cases even $S$ is not smooth enough, in which case one has to use higher orders of WKB and/or to match it with other types of approximate solutions in the regions where they are available~\cite{kessler2007extinction}. However, in most cases the controlling-factor analysis is adequate~\cite{assaf2017wkb}, and our problem is no exception.

  \item The \textbf{two-destination approximation}, under which we replace the original process with an effective walk that has the same mean and variance, facilitates the utilization of the WKB method tremendously and we use it throughout this manuscript. In this paper we have employed this approximation in a specific way: the chance to jump to the left or to the right is chosen to be $1/2$, while the jump lengths are made asymmetric, either $r - x + \sqrt{\mathrm{Var}[m]/N^2}$ or $r - x - \sqrt{\mathrm{Var}[m]/N^2}$ [with $r$ as defined in Eq.~(\ref{eq01})]. Importantly, this choice is by no means necessary: any combination of chances and jump lengths that keeps the mean and the variance fixed is acceptable.

      This particular implementation of the two-destination approximation (equal jump probabilities, different jump lengths) may fail. In particular, if both the destinations are to the left of the original location $x=n/N$, then Eq.~(\ref{eq10}) allows only one solution, $q=0$, because a two-destination walk of this type cannot end up in fixation. Similarly, when both the destinations are to the right of the original location, then the approximated walk does not allow extinction. The actual process [Eq.~(\ref{WKB1})] still has a small chance of extinction or fixation, since it supports larger (although improbable) jumps. As a result, when $\sqrt{x(1-x)/N}$ becomes too close to $r-x$, the two-destination approximation that we have employed here fails while Eq.~(\ref{WKB1}) is still valid.

	Nevertheless, the failure of Eq.~(\ref{eq10}) in the case when $\sqrt{\mathrm{Var}[m]/N^2} < |r|$ springs from the particular choice of the two destinations that we have made, and is not a limitation of the general approach itself of replacing the full many-destination process with a two-destination one. Although we do not do so in this paper, it is possible to choose the two destinations differently, with a corresponding careful choice of the probabilities, such that the mean and the variance of the steps are preserved even when $\sqrt{\mathrm{Var}[m]/N^2} < |r|$, and an evolution of the system in both directions, fixation and extinction, is allowed.

      Interestingly, in the case of fixed selection, the two-destination approximation as employed here \emph{always} fails in the large-$N$ limit. In the middle regime, where both $x$ and $1-x$ are ${\cal O} (1)$, $r(x)-x$ is an $N$-independent number while the demographic term $\sqrt{x(1-x)/N}$ goes to zero. However, under fixed selection (as opposed to the cases of fluctuating selection considered below) $\Pi(x)$ is almost fixed in this middle regime where $\sqrt{x(1-x)/N} \ll 1$.  In the case of a beneficial mutant ($s>0$), $\Pi$ grows and reaches values that are very close to one in the inner regime $x \ll 1$ (where $r(x)-x$ is ${\cal O} (1/N)$, so it is comparable with $\sqrt{x/N}$), while for a deleterious mutant $\Pi$ is almost zero until the outer regime $1-x \ll 1$. Therefore, the technical failure of the two-destination WKB approximation in the middle regime has no effect on the outcome.

  \item The \textbf{small-$q$ approximation}, or more exactly small $qsx(1-x)$ and small $q\sqrt{x(1-x)/N}$, was used to derive Eq.~(\ref{eq11}) from Eq.~(\ref{eq10}). In the following sections, where the effect of a fluctuating environment is discussed, we obtain analogous equations that have the same form as Eq.~(\ref{eq10}), for which the arguments of the exponent and cosh functions are not small. For these cases we present, in the next section, alternate approximations.

\end{itemize}

\section{The fundamental transcendental equation} \label{fundamental}

The first-order WKB method, when employed using the two-destination approximation as described in the last section, yields naturally a transcendental equation of the form
\begin{equation} \label{mtrans}
e^{q s_\text{e}}\cosh(q \sigma_\text{e}) = 1.
\end{equation}
In this equation $s_\text{e}$ is the effective  bias of the population towards fixation or towards extinction, while $\sigma_\text{e}$ is the effective stochasticity (demographic plus environmental). The value of $q$ depends strongly (see below) on the ratio between $s_\text{e}$ and $\sigma_\text{e}$.  Equation~(\ref{mtrans}) has no known closed-form solution in terms of elementary functions. In this section we present approximate solutions that work well in various sectors.

First, we write Eq.~(\ref{mtrans}) as
\begin{equation} \label{ap21}
\ln \cosh (q \sigma_\text{e}) = -q s_\text{e}.
\end{equation}
Because the $\cosh$ function is symmetric, Eq.~(\ref{ap21}) reveals that $q(-s_\text{e}) = -q(s_\text{e})$.

Second, with the definitions $\tilde{q} \equiv q \sigma_\text{e}$  and $\tilde{s} \equiv s_\text{e}/\sigma_\text{e}$, we obtain the one-parameter equation
\begin{equation} \label{eq20}
e^{\tilde{q}\tilde{s}}\cosh(\tilde{q}) = 1.
\end{equation}
Now we can identify three different sectors, and fit an approximate solution in each of them separately.

\begin{figure}[h]
\begin{center}
\includegraphics[width=10cm]{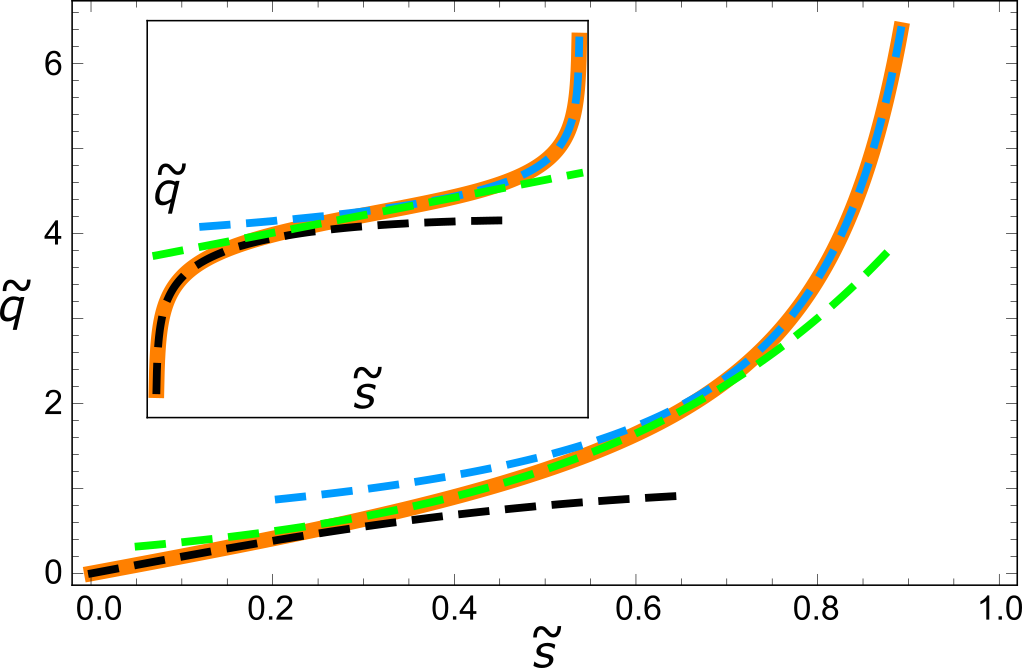}
\end{center}
\vspace{-0.5 cm}
\caption{Numerical solution (solid orange curve) of the transcendental equation (\ref{eq20}) as a function of $\tilde{s}$, together with the small-$\tilde{q}$ approximation (Eq.~(\ref{lowq}), black dashed curve), the middle-$\tilde{q}$ approximation (Eq.~(\ref{midq}), green dashed curve) and the large-$\tilde{q}$ approximation (Eq.~(\ref{largeq}), blue dashed curve). A linear scale is used in the main panel, and the same plot is shown using a semi-logarithmic scale in the inset. Note the linear nature of the green curve in the inset. \label{trans}}
\end{figure}

\begin{enumerate}
  \item \textbf{In the small-$\tilde{q}$ sector}, one may expand the exponent and cosh functions in Eq.~(\ref{eq20}) to the second order in $\tilde{q}$. This yields
      \begin{equation}
      \tilde{q} = -\frac{2\tilde{s}}{\tilde{s}^2+1},
      \end{equation}
       or
      \begin{equation} \label{lowq}
       q = -\frac{2s_\text{e}}{s_\text{e}^2+\sigma_\text{e}^2}.
       \end{equation}

Since the DA assumes that $\Pi$ is smooth, when this approximation is applicable $S = \ln \Pi$ is clearly smooth as well, so $q = S'$ has to be small. Accordingly, one expects that when the diffusion approximation works, the system is in the small-$\tilde{q}$ sector. In fact, the diffusion approximation in its canonical form (\ref{canonical}) suggests that the parameter $2 s_\text{e}/\sigma_\text{e}^2$ controls the results, so it corresponds to the subsector $|s_\text{e}| \ll \sigma_\text{e}$ of the small-$\tilde{q}$ sector.

  \item \textbf{In the large-$\tilde{q}$ sector}, the $\cosh(\tilde{q})$ factor in Eq.~(\ref{eq20}) may be replaced by an exponent, $\cosh(\tilde{q}) \approx \exp(|\tilde{q}|)/2$. This leads to $\tilde{q}\tilde{s_\text{e}} + |\tilde{q}| = \ln 2$, so
   \begin{equation}
   \tilde{q} = \frac{\ln 2}{\tilde{s} - {\rm sign}(s_\text{e})},
   \end{equation}
      and
  \begin{equation}  \label{largeq}
  q = \frac{\ln 2}{s_\text{e}-{\rm sign}(s_\text{e}) \sigma_\text{e}}.
  \end{equation}

As $\sigma_\text{e}$ approaches $|s_\text{e}|$ from above, the moves that are against the current (i.e., backward moves if $s_\text{e}>0$ and forward ones if $s_\text{e}<0$) become rarer and $q$ increases. Formally, Eq.~(\ref{eq20}) implies that $q$ diverges at $\sigma_\text{e} = |s_\text{e}|$, but this divergence does not reflect a discontinuous change in the chance of fixation. Instead (as discussed in section \ref{outlook}) it reflects the breakdown of the two-destination approximation. Once $\sigma_\text{e} \le |s_\text{e}|$, moves against the deterministic current occur only through (rare) events at the tails of the jump distribution, so one cannot approximate the whole spectrum by its standard deviation. Therefore, the analysis suggested here becomes less reliable when the system enters deep inside the large-$\tilde{q}$ sector.

  \item  \textbf{In the intermediate-$\tilde{q}$ sector}, none of the above approximations work. When $\tilde{q}$ is plotted against $\tilde{s}$ on a semi-logarithmic scale, one finds that the small-$\tilde{q}$ approximation works well for $\tilde{s}<0.25$ and the large-$\tilde{q}$ solution works well for $\tilde{s}>0.7$. Fortunately, in the intervening region $ 0.25 < \tilde{s} < 0.7$, $\ln(\tilde{q})$ is to a good approximation a straight line in $\tilde{s}$ (see the inset panel in Figure~\ref{trans}). Fitting a straight line for $\ln(\tilde{q})$ in this intermediate-$\tilde{q}$ sector provides us with the approximation
      \begin{equation}
      \tilde{q} = -{\rm sign} (s_\text{e}) e^{3|\tilde{s}|-1.3}.
      \end{equation}
      As a result,
  \begin{equation}  \label{midq}
  q = -{\rm sign} (s_\text{e}) \frac{e^{3 |s_\text{e}|/\sigma_\text{e}-1.3}}{\sigma_\text{e}}.
  \end{equation}
\end{enumerate}

Figure~\ref{trans} shows how these three solutions cover the whole $\tilde{s}$-range. Table \ref{table1} summarizes the $q(s_\text{e},\sigma_\text{e})$ relationships used hereon as approximate solutions of the fundamental transcendental equation.

\begin{table}[h]
\caption{}
\centering
\begin{tabular}{|c | c |c|}
\hline\hline
 small $q$ &  $|\tilde{s}| < 0.25$ & $q \approx -\frac{2 s_\text{e}}{(s_\text{e}^2+\sigma_\text{e}^2)}$  \\ \hline
intermediate $q$ & $0.25< |\tilde{s}| < 0.7$ & $q \approx -{\rm sign}(s_\text{e}) \frac{e^{3 (s_\text{e}/\sigma_\text{e})-1.3}}{\sigma_\text{e}}$ \\ \hline
 large $q$ & $0.7<|\tilde{s}|$   &  $q \approx \frac{\ln 2}{s_\text{e} - {\rm sign}(s_\text{e})\sigma_\text{e}}$  \\ \hline \hline
\end{tabular}
\label{table1}
\end{table}

\section{Chance of ultimate fixation under fluctuating selection: the model and its direct numerical solution} \label{model1}

Now we consider the case of a fluctuating environment. The variability in the environment manifests itself in a corresponding variability of the selection parameter. As before, we assume that the full community has a population of size $N$, with $n$ mutant-type individuals and $N-n$ wild-type individuals. Now the state of the system is fully characterized by $N$, $n$ and $k$, the state of the environment. These quantities dictate the transition probabilities $W_{\{n,k\} \to \{n+m,k'\}}$, where $m$ is the change in the abundance of the mutant type during a time-step (and may be negative). The chance of ultimate fixation, $\Pi_n^k$, satisfies the BKE,
\begin{equation}  \label{eqn1}
\Pi_n^k = \sum_{m,k'} W_{\{n,k\} \to \{n+m,k'\}} \Pi_{n+m}^{k'},
\end{equation}
with the boundary conditions $\Pi_0^k = 0 $ and $\Pi_N^k=1$. As in the fixed selection case, since $\sum_{m,k'} W_{\{n,k\} \to \{n+m,k'\}} =1$, any constant $C_1$ is a solution of (\ref{eqn1}) and the general solution is the linear combination $\Pi_n^k = C_1 + C_2 \overline{ \Pi}_n^k$, where $\overline{ \Pi}_n^k$ is a particular non-trivial (non-constant) solution of (\ref{eqn1}) and $C_1$ and $C_2$ are determined by the absorbing boundary conditions.

The diffusion approximation approach assumes that $\Pi$ is smooth over its state space~\cite{crow1970introduction,ewens2012mathematical,takahata1975effect,takahata1979genetic,danino2018fixation,meyer2018noise}. In systems with fluctuating selection, the DA requires $\Pi$ to be smooth also over the different environmental states $k$, for instance with the condition $\Pi_n^{k_1} - \Pi_n^{k_2} \ll 1$, for two environmental states $k_1$ and $k_2$~\cite{danino2018stability,danino2018stability,danino2018fixation}. This requirement is not satisfied when the persistence time of the environment is comparable to the fixation time~\cite{mustonen2008molecular,cvijovic2015fate}; in what follows we assume that $N$ is large enough such that this condition is satisfied, which allows us to deal only with the non-smoothness of $\Pi$ over the population states, and to average all $\Pi_n^k$ to one $\Pi_n$.

The dynamics used in Section~\ref{fixed} are extended to include fluctuating selection, so $s$ is now picked at random in each generation. To facilitate the calculations we assume dichotomous stochasticity, with $s(t)$ either $s_0+\sigma$ or $s_0-\sigma$, both with probability $1/2$.
This is  a useful minimal model for fluctuating selection (see \cite{yahalom2019comprehensive}, Appendix~A) and one can quite easily extend the procedure presented here to similar scenarios.

The great advantage of this discrete time model, in which a new environment is independently picked in every step, is its amenability to direct numerical solution. In general the dimension of a Markov matrix for a model with fluctuating selection is $NK \times NK$, where $N$ is the number of individuals and $K$ is the number of environmental (selection) states. Here, because the environment is picked at random in each generation with no correlations, one can average over all $s$ states to obtain an $N \times N$-dimensional matrix.

This task becomes even simpler when the noise is dichotomous. The chance of the mutant type to capture any given slot out of the $N$ open slots in the next haploid generation is
\begin{equation}
r_{\pm} = \frac{n e^{s_0 \pm \sigma}}{n e^{s_0 \pm \sigma} + (N-n)},
\end{equation}
and this leads to a simple form for the transition probabilities,
\begin{equation} \label{destinations}
W_{n \to n+m} = \frac{1}{2} \binom{N}{n+m} \left[ r_+^{n+m}(1-r_+)^{N-n-m}+ r_-^{n+m}(1-r_-)^{N-n-m} \right].
\end{equation}

Therefore, the BKE takes the form $\Pi_n = {\cal W}_{n,n+m} \Pi_{n+m}$, where ${\cal W}$ is the $(N-1) \times (N-1)$ Markov matrix whose $(n,n+m)$-th element is given in (\ref{destinations}).
One can solve this BKE with its boundary condition by inverting the Markov matrix ${\cal W}$ (minus the identity matrix $\mathbb{1}$) and multiplying it by $f_n$,
\begin{equation} \label{exactnum}
\Pi = -({\cal W}-\mathbb{1})^{-1} f,
\end{equation}
where the elements of the vector
\begin{equation}
f_n =  W_{n \to N}
\end{equation}
are the chance of fixation in the next generation when the current mutant population is $n$. Throughout this paper we have employed this procedure to find direct numerical solutions for $\Pi_n$, which are then compared with the outcomes of the various approximation techniques used.

\section{The WKB approach and a scalable numerical technique}\label{s7}

The direct numerical solution presented in the last section requires matrix inversion whose numerical complexity is ${\cal O} (N^3)$. In this section we present the WKB approach for a system with fluctuating selection and adopt it as an alternative numerical technique. This new technique is scalable: it requires (up to) five independent steps of one-dimensional numerical integrations, so the numerical effort is essentially $N$-independent. Besides, the discussion here lays the groundwork for the derivation (in the next sections) of analytical solutions by combining the WKB with the asymptotic matching technique.

\subsection{Fluctuating selection in logit space}

For the sake of convenience, we define a new parameter, the logit function $z$, whose relationship with the frequency $x$ is given by
\begin{equation} \label{logit1}
z \equiv \ln \left(\frac{x}{1-x}\right), \qquad \text{so} \qquad x \equiv \frac{e^z}{e^z+1}.
\end{equation}
Since $z$ and $x$ are simple functions of each other, we will switch between them often and in some cases will present mixed expressions that utilize both the variables. One can easily translate these expressions to those containing only $x$ or only $z$ using Eq.~(\ref{logit1}).

We define the effective selection, $s_\text{e}(z)$, and the effective stochasticity, $\sigma_\text{e}(z)$, as the mean and the standard deviation of $\Delta z$ (the change in $z$) per generation. The two-destination approximation is used to model the dynamics as a simple effective walk, where all the effects of selection and stochasticity (including demographic stochasticity) manifest themselves in the lengths of the possible jumps. In the version adopted here, the allowed jumps are $\Delta z = s_\text{e}+\sigma_\text{e}$ and $\Delta z = s_\text{e}-\sigma_\text{e}$, each with probability $1/2$. Such a walk keeps the values of the mean and the variance of $\Delta z$ the same as their values in the original process.

For a \emph{given} value of $s$, the  $\Delta z$ destinations are calculated as follows. First we find
\begin{align}
x &\to \frac{xe^s}{1-x+xe^s} \pm \sqrt{\frac{x(1-x)}{N}} = \frac{xe^s}{1-x+xe^s} \left(1 \pm \sqrt{\frac{A^2 (1-x)}{Nxe^{2s}}} \right), \nonumber \\
1-x &\to \frac{1-x}{1-x+xe^s} \mp \sqrt{\frac{x(1-x)}{N}} = \frac{1-x}{1-x+xe^s} \left(1 \mp \sqrt{\frac{A^2 x}{N(1-x)}} \right),
\end{align}
where
\begin{equation}
A \equiv 1-x+xe^s = \frac{1+e^{s+z}}{1+e^z}.
\end{equation}

If $N$ is taken to be a large parameter (larger than any other parameter in the problem, like $A$ and so on), then to the leading order in $N$ we have
\begin{equation}
z \to z' = z+s + \ln\left(1 \pm  \sqrt{\frac{A^2 (1-x)}{Nxe^{2s}}}\right) - \ln\left(1 \mp \sqrt{\frac{A^2 x}{N(1-x)}} \right) \approx z+s \pm \left(  \sqrt{\frac{A^2 (1-x)}{Nxe^{2s}}} +  \sqrt{\frac{A^2 x}{N(1-x)}} \right).
\end{equation}
This simplifies to
\begin{equation} \label{defB}
z  \to z'  \approx z+s \pm B(s),
\end{equation}
where
\begin{equation}\label{Bdef}
B(s) \equiv  \frac{1+\cosh(s+z)}{\sqrt{N} \cosh(z/2)}.
\end{equation}

Now let us consider the case of fluctuating selection. Since $s$ (for dichotomous noise) takes two values, $s_0+\sigma$ and $s_0-\sigma$, the four-destination BKE (equivalent to Eq.~(\ref{WKB2}) above) is
\begin{equation}
\Pi(z) = \frac{1}{4} \sum_{\substack{\zeta_1 =-1,1 \\ \zeta_2 = -1,1}}  \Pi \left[ s_0 + \zeta_1 \sigma +  \zeta_2  B( s_0 + \zeta_1 \sigma) \right].
\end{equation}
Accordingly,
\begin{equation} \label{se}
s_\text{e}(z) = \mathbb{E}[\Delta z] = s_0,
\end{equation}
and
\begin{equation} \label{sige}
\sigma_\text{e}^2 (z)= \frac{1}{4} \left\{\sum_{\substack{\zeta_1 = -1,1 \\ \zeta_2 = -1,1}} \left[ s_0 + \zeta_1 \sigma + \zeta_2 B( s_0 + \zeta_1 \sigma) \right]^2\right\} - s_0^2.
\end{equation}

From now on, we will employ an effective two-destination approximation, where $z \to z + s_\text{e} \pm \sigma_\text{e}$. As a result, the fundamental transcendental equation takes the form
\begin{equation} \label{WKB5}
e^{q(z) s_\text{e}(z)} \cosh[q(z) \sigma_\text{e}(z)] = 1.
\end{equation}
Since we will solve the equation in terms of the logit variable $z$, the special solution  $\overline{\Pi}$  will be $\overline{\Pi} = e^{S(z)}$ where $S(z) = \int q(z) dz$ and $q(z)$ is the non-trivial solution of (\ref{WKB5}). In what follows we will use in some cases the solution $q(n)=q(xN)$, and in these cases a change of variables in the integral yields  $S(n) = \int [q(n)/n] dn$.

\subsection{A scalable numerical solution}\label{subsec:scalable}

Given Eq.~(\ref{WKB5}) one may adopt a conceptually simple numerical procedure to find $S_n$ and $\Pi_n$. First one solves (\ref{WKB5}) to obtain $q(z)$ for each value of $z_n  = \ln[n/(N-n)]$, from the minimal point $z_\text{min} \approx -\ln N$ to the maximal point $z_\text{max} \approx \ln N$. Summing $q$ over these $z$-values yields $S_n$ and the special solution for $\Pi_n$, and imposition of the boundary conditions yields the full solution.

However, the use of this technique necessitates a solution of $N$ transcendental equations of the form  (\ref{WKB5}). Since zero is always a solution and we need the other, non-trivial solution, this task is complicated for a single $n$-value, and solving $N$ such equations requires considerable numerical effort.

\begin{figure}[h]
\begin{center}
\includegraphics[width=18cm]{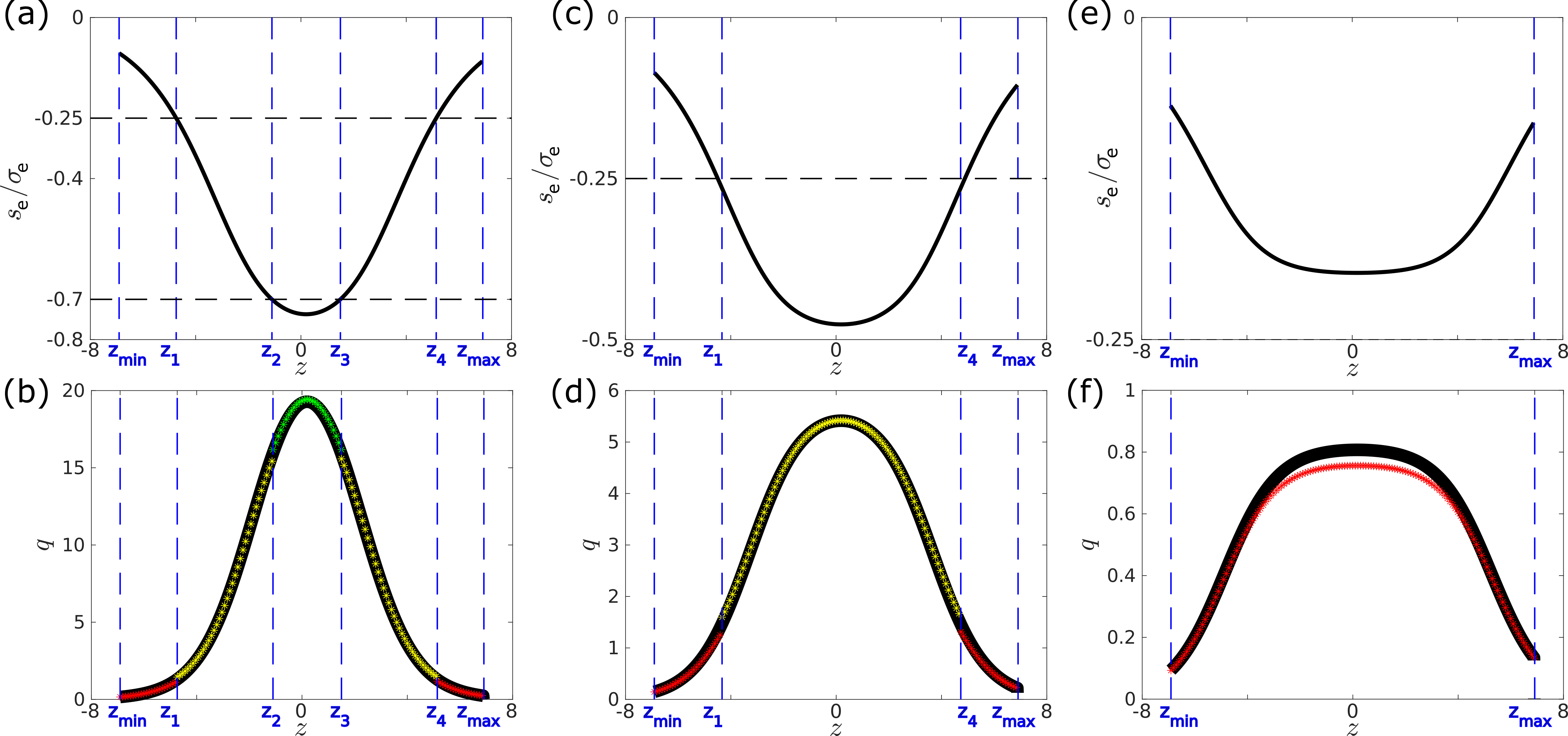}
\end{center}
\vspace{-0.5 cm}
\caption{The different panels in this figure show the values of $\tilde{s} = s_\text{e}/\sigma_\text{e}$ (top panels), and the corresponding values of  $q$ (bottom panels), as functions of $z$. All quantities are plotted against $z \equiv \ln[x/(1-x)]$, between $z_\text{min} \approx -\ln N$ (one individual) and $z_\text{max} \approx \ln N$ ($N-1$ individuals). $z_1$ and $z_4$ are defined as the $z$-values for which $|s_\text{e}/\sigma_\text{e}| = 0.25$, and $z_2$ and $z_3$ are the $z$-values for which $|s_\text{e}/\sigma_\text{e}| = 0.7$. As explained in the main text, $0.25$ and $0.7$ form two threshold values of $\tilde{s}$. When the maximum absolute value of $\tilde{s}$ is larger than $0.7$ [panel (a)] then there are five $z$-regions, two corresponding to the small-$q$ sector, two to the middle-$q$ sector and one to the large-$q$ one, as demonstrated in panel (b). When the maximum absolute value of $\tilde{s}$ is between $0.2$5 and $0.7$ [panel (c)] then there are two $q$-sectors (small and middle) and three $z$-regions [panel (d)], while if the maximum absolute value of $\tilde{s}$ is less than $0.25$ [panel (e)] then there is only one $q$-sector [panel(f)]. In all the bottom panels, red, yellow and green asterisks denote the approximate analytical expressions for small $q$, intermediate $q$ and large $q$, respectively, as given in Table~\ref{table1}, while the solid black lines show $q$-values as extracted from the numerical solution of Eq.~(\ref{WKB5}). The parameters are: $s_0 = -0.1$; $N = 1000$; and  $\sigma = 0.12$ [panels (a)-(b)], $0.2$ [panels (c)-(d)], and $0.5$ [panels(e)-(f)]. \label{fig444}}
\end{figure}

To overcome this difficulty, we utilize the approximate solutions presented in Section~\ref{fundamental}. Once $\sigma_\text{e}(z)$ and $s_\text{e}(z)$ are given, via Eqs.~(\ref{se}) and (\ref{sige}), the value of $\tilde{s} = s_\text{e}/\sigma_\text{e}$ suggests a solution obtainable from Table~\ref{table1}. In Figure~\ref{fig444} we present a few typical profiles of $q(z)$ versus $z$, between $z_\text{min}$ and $z_\text{max}$. In all cases $q(z)$ is small close to the extinction/fixation points, since the demographic noise is strong in these regions so $\sigma_\text{e}$ increases while $s_\text{e}$ is kept fixed. In some cases the maximum value of $q$ lies in the large-$q$ sector [Fig.~\ref{fig444} (a)-(b)], in some cases it lies in the intermediate-$q$ sector [Fig.~\ref{fig444} (c)-(d)], and in some cases it lies in the small-$q$ sector [Fig.~\ref{fig444} (e)-(f)].

Our scalable numerical technique is easy to use. We will describe it for the most complicated case where $q(z)$ reaches the large-$q$ sector. Its adaptation to the other cases follows immediately.

First we identify the different regimes by solving for $z_1$ and $z_4$, the two points at which $|\tilde{s}(z)| = 0.25$, and for $z_2(x)$ and $z_3(x)$, the two points at which $|\tilde{s}(z)| = 0.7$. In each regime we use the appropriate expression for $q(z)$ (note that in our case $s_\text{e} = s_0$, while $\sigma_e$ depends on $z$),
\begin{eqnarray}\label{eq:M_regions}
z_\text{min} \le z \le  z_1: \qquad q_1(z) &=& \frac{-2s_0}{s_0^2 + \sigma_\text{e}^2}, \nonumber \\
z_1 \le z \le z_2: \qquad q_2(z) &=& -{\rm sign} (s_0) \frac{e^{3 |s_0|/\sigma_\text{e}-1.3}}{\sigma_\text{e}}, \nonumber \\
z_2 \le z \le z_3: \qquad q_3(z) &=& \frac{\ln 2}{s_0-{\rm sign}(s_0) \sigma_\text{e}},  \\
z_3 \le z \le z_4: \qquad q_4(z) &=& -{\rm sign} (s_0) \frac{e^{3 |s_0|/\sigma_\text{e}-1.3}}{\sigma_\text{e}}, \text{ and}\nonumber \\
z_4 \le z \le  z_\text{max}: \qquad q_5(z) &=& \frac{-2s_0}{s_0^2 + \sigma_\text{e}^2}. \nonumber
\end{eqnarray}
This scheme, with five regions, is required only if the maximum value of $\tilde{s}$ is larger than $0.7$. Otherwise, if the maximum value is between $0.25$ and $0.7$ then only three regions are required, and if the maximum value is smaller than $0.25$ then only one region is needed.

Once the values of $z_1$ to $z_4$ are found, the quantities $I_1$ to $I_5$ are defined as the integrals over $q(z)$ in the relevant regimes,
\begin{equation}
I_1 \equiv \int_{-\infty}^{z_1} q_1(z) dz, \qquad I_2 \equiv \int_{z_1}^{z_2} q_2(z) dz, \quad \text{etc.}
\end{equation}
In our numerical integrations we replaced $-\infty$ by $\kappa z_\text{min}$, where $\kappa$ is some large number, and correspondingly $+\infty$, at the right end of the integration that yields $I_5$, by $\kappa z_\text{max}$.

For any given value of $z$, $S(z)$ is the integral over $q(z)$ from $-\infty$ to $z$. To be precise,
\begin{eqnarray}
\text{if }  z \le z_1: \qquad S(z) &=& \int_{-\infty}^{z} q_1(z) dz, \nonumber \\
\text{if }  z_1 \le z \le z_2: \qquad S(z) &=& I_1 + \int_{z_1}^{z} q_2(z) dz, \\
\text{if }  z_2 \le z \le z_3: \qquad S(z) &=& I_1 + I_2 + \int_{z_2}^{z} q_3(z) dz, \qquad \text{... and so on.} \nonumber
\end{eqnarray}
Finally, the chance of ultimate fixation is given by
\begin{equation}
\Pi(z(x)) = \frac{e^{S(z)}-1}{e^{\sum  I_j}-1}.
\end{equation}
In the denominator above, the sum is taken over the total number of $I_j$ present [depending on the number of regions in (\ref{eq:M_regions})].

\begin{figure}[h]
\begin{center}
\includegraphics[width=8.5cm]{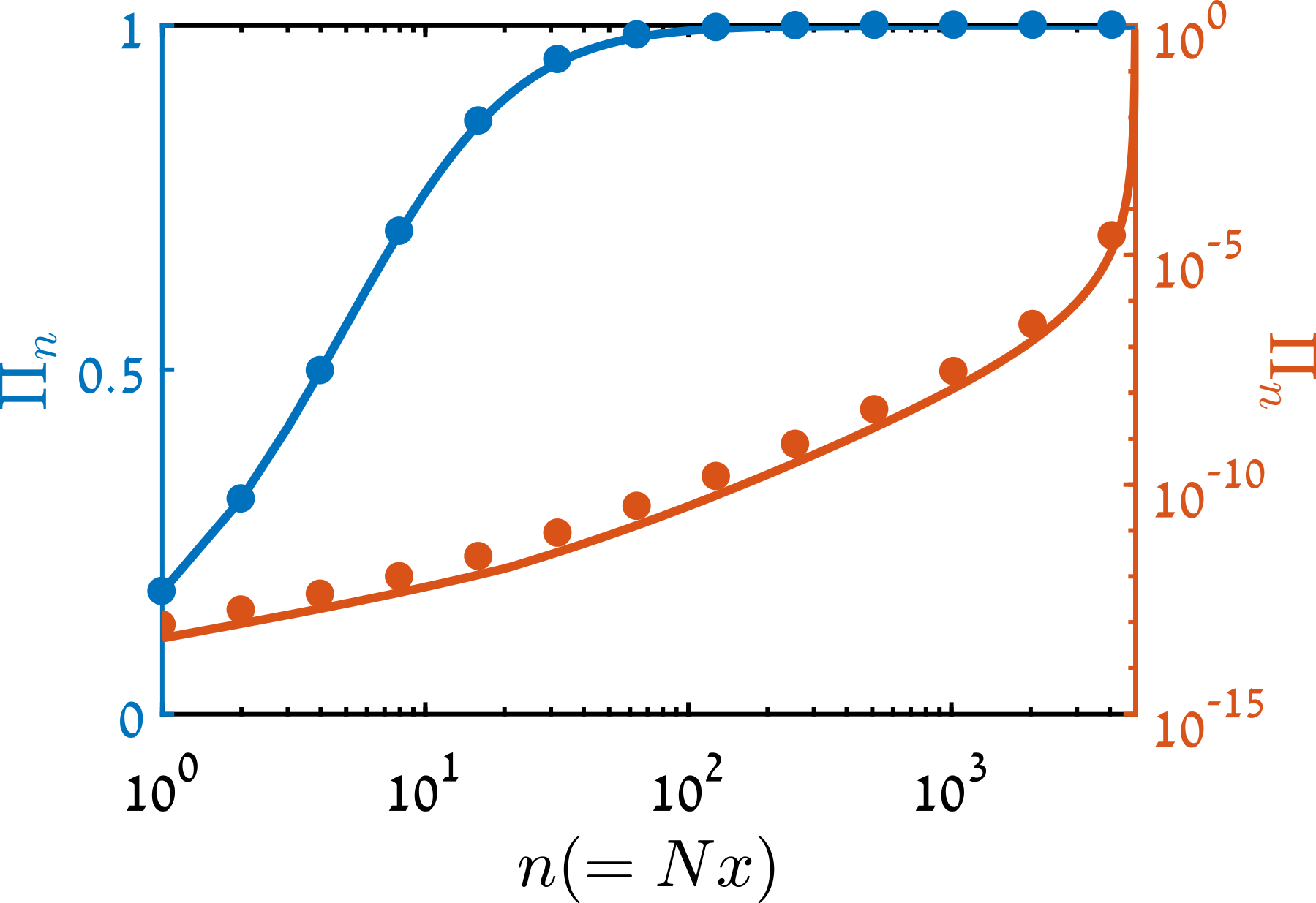}
\end{center}
\vspace{-0.5 cm}
\caption{The effectiveness of the scalable numerical technique, as described in section~\ref{subsec:scalable}, for a system with fluctuating selection. The chance of ultimate fixation $\Pi_n$ is plotted against $n$ for a beneficial (blue, left $y$-axis, linear scale) and deleterious (orange, right $y$-axis, logarithmic scale) mutant. Circles represent the results of the direct numerical solution of the BKE, Eq.~(\ref{exactnum}), while the solid lines show the scalable numerical solutions with $\kappa = 10$. The parameters are $N = 5000$, $\sigma = 0.3$ and $s_0 =  \pm 0.1$. As discussed in the text, the diffusion approximation breaks down well below $\sigma = 0.3$. \label{fig555}}
\end{figure}

All in all, for our system this technique involves the numerical solution of at most five transcendental equations of the form $\tilde{s} = 0.25$ (or $0.7$) and at most five numerical integrations over known functions. Typical results obtained using this scalable numerical technique are compared with the direct numerical (matrix inversion) results in Figure~\ref{fig555}.

In our system the profiles of $s_\text{e}$ and $\sigma_\text{e}$ are rather simple: $s_\text{e}$ is $z$-independent and $\sigma_\text{e}$ grows from the edges and reaches a maximum around $z=0$. In other scenarios the function $q(z)$ may be more complicated, and consequently there may be more than five regions in the $q$-$z$ plot that need to be considered separately in order to employ the numerical technique described here. Nevertheless the basic method holds good and remains independent of the size $N$ of the community, unlike the numerical solution of the full dynamics.

There exist other numerical approaches (for high dimensions) in the literature, which depend on the identification of an optimal path between two fixed points of the deterministic dynamics~\cite{lindley2013iterative}. When transitions between these fixed points (through stochastic tunneling) depend on rare events like an improbable accumulation of bad years, the probability of such an optimal path determines the chance of transition. Our method has two advantages with respect to optimal-path techniques. First, as explained in the following sections, our technique does not rely on the existence of a unique optimal path and works well even when extinction (or fixation) is not a rare event. Second, with our method one may calculate fixation probabilities and transition times between any two points in the phase space, including stochasticity-induced fixed points like those that emerge in Chesson's lottery model (see~\cite{dean2020stochasticity}). Such fixed points are by definition invisible to the deterministic dynamics and hence cannot be treated by the optimal-path techniques, whereas our approach, which simply extends the diffusion approximation by relaxing its smoothness requirement, is applicable in such scenarios as well.

\section{Asymptotic matching} \label{s3}

With the WKB procedure set up, the set of integrals defined in the last section, if solvable, provides an analytical expression for the chance of ultimate fixation. In this and in the next section we consider cases in which such an analytical expression is attainable. In general, this depends on the existence of a middle regime where the effect of demographic stochasticity is negligible. Such a regime always exists if $N$ is large enough and $\sigma > |s_0|$. When the system supports a middle regime, one may employ (as in~\cite{danino2018fixation,meyer2018noise}) a standard asymptotic matching procedure, where the solutions in the inner and the outer regimes are matched with the middle solution in the large-$N$ limit. Here we first quantify the conditions under which a middle regime exists, and then re-derive the DA results obtained in~\cite{danino2018fixation,meyer2018noise} for a model with fluctuating selection. In the succeeding sections we explain how to derive the same expressions and much better approximations using the WKB method.

\subsection{The middle regime at large $N$} \label{middle5}

We define the middle regime as the range of $z$ in which demographic stochasticity (genetic drift) is negligible. On examining Eqs.~(\ref{Bdef})-(\ref{WKB5}) one sees that demographic noise affects the outcome of the calculations only via the $B$ term in Eq.~(\ref{sige}). Since the $B$ term scales like $1/\sqrt{N}$, for any fixed $z$ it eventually (as $N$ increases) becomes negligible when $B(s_0 \pm \sigma) \ll s_0 \pm \sigma$. Therefore, when $q(x)$ is bounded and $N \to \infty$, the system admits a middle regime.

Let us calculate the critical value of $N$, above which one should expect a finite $z$-region in which demographic stochasticity is negligible. On minimizing the expression for $B$ in Eq.~(\ref{Bdef}) with respect to $z$ and discarding those $z$-values that result in complex or negative values for $B$, we find that the minimal value of $B$ appears at
\begin{equation}
z^* = 2 \cosh ^{-1}\left(\frac{1}{2} \sqrt{\sqrt{9 \cosh^2(s)+2 \cosh (s)-7}+3 \cosh (s)-1}\right),
\end{equation}
where $s$ can take the values $s_0 \pm \sigma$. So we have
\begin{equation}
B(z^*) = \frac{2+7s^2/2}{\sqrt{N}} + {\cal O}(s^4) \approx \frac{2}{\sqrt{N}}.
\end{equation}
Therefore, when
\begin{equation} \label{cond3}
 \sqrt{N} \gg \frac{2}{\sigma-|s_0|},
\end{equation}
the value of $q$ in the middle zone (around $z^*$) is independent of the strength of demographic stochasticity. Since $s_0$ and $\sigma$ are $n$-independent, this implies that the value of $q$ in this middle zone does not depend on $z$ (or $x$ or $n$). For each system that satisfies (\ref{cond3}) one expects a plateau in the $q$-$z$ diagram. For instance, in Fig.~\ref{fig444} panel (b), $\sqrt{N}(\sigma-|s_0|)/2 = 0.316$ and there isn't a plateau, whereas in Fig.~\ref{fig444} panel (f), $\sqrt{N}(\sigma-|s_0|)/2 = 6.32$ and the plateau is seen. Note that the value of $N$ above which the plateau appears scales like $4/(\sigma - |s_0|)^2$ and so it diverges as $\sigma \to |s_0|$.

The existence of a middle regime for large $N$ suggests an \textbf{asymptotic matching approach}: one can solve for the relevant quantity in the middle regime and match it to the solutions in the inner ($x \ll 1$) and the outer ($1-x \ll 1$) regimes, where the system is close to fixation/extinction and the demographic noise dominates its behavior. In the next subsection we review the results obtained using the diffusion approximation, and then we show how to derive the same, and improved, results using WKB.

\subsection{Three frequency regimes and asymptotic matching: the diffusion approximation} \label{3f}

In former studies \cite{danino2018fixation,meyer2018noise} a model with fluctuating selection has been solved by employing the diffusion approximation. In these studies, $\Pi_{n+m}$ was expanded to second order in $m$ (as in Eq.~(\ref{smooth}) above) and the relevant terms were collected to yield a differential equation for the chance of ultimate fixation as a function of $x$.

These works considered a Moran model, where the environment flips erratically between the two states $s_0 \pm \sigma$ and the persistence times are picked from an exponential distribution whose mean is $\tau$ generations, where a generation is defined as $N$ elementary birth-death events. In their regime of validity (namely, when $s_0 \ll \sigma \ll 1$), these results hold also for our model, provided that one takes $\tau=1$ and replaces $N$ by $2N$. (The strength of demographic stochasticity is $V/N$, where $V$ is the variance of the number of offspring per individual during its lifetime. In Moran models the distribution of offspring per individual is geometric so $V=2$, while in Wright-Fisher models the distribution is Poissonian and $V=1$. So to match Wright-Fisher results with a Moran model, the effective population size has to be doubled.)

Translating the expressions of \cite{danino2018fixation,meyer2018noise} to the Wright-Fisher language, the chance of ultimate fixation satisfies
\begin{equation}\label{eq30}
\left[\frac{1}{N} +  \sigma^2 x (1-x) \right] \Pi'' + \left[2s_0  +  \sigma^2 (1-2x)\right]\Pi' = 0.
\end{equation}
Note that in the limit $\sigma \to 0$, Eq.~(\ref{eq30}) converges to Eq.~(\ref{eq5}). The boundary conditions are, as before, $\Pi(x=0) = 0$ and $\Pi(x=1) = 1$.

Eq.~(\ref{eq30}) is a first-order linear differential equation and may be solved directly using an integrating factor, but the outcome is messy and hard to interpret. The strategy adopted in \cite{danino2018fixation,meyer2018noise} is based instead on asymptotic matching. The segment $0 \le x \le 1$ is divided into the following three parts:
\begin{itemize}
  \item The inner regime (close to extinction), where $x \ll 1$ and $1-x \approx 1$.
  \item The middle regime, where $x(1-x) \gg 1/N$. The demographic stochasticity term $1/N$ in (\ref{eq30}) may be neglected in this regime. Unlike (\ref{cond3}), here the width of the middle regime scales with $1/\sigma^2$, with no $s_0$-dependence. This happens since the treatment in \cite{danino2018fixation,meyer2018noise} assumes $s_0 \ll \sigma$.
  \item The outer regime (close to fixation) where $1-x \ll 1$ and $x \approx 1$.
\end{itemize}
In each of these regimes Eq.~(\ref{eq30}) simplifies and may be solved explicitly, with solutions of the form $C_i + C_j \bar{\Pi}(x)$, where $C_i$ and $C_j$ are constants (for explicit solutions see the next section). These constants are determined by the requirement that the different solutions match each other when their regions of validity overlap. The inner solution matches the middle one when both $x \ll 1$ and $x \gg 1/N$ (i.e., when $Nx \gg 1$ but $x \ll 1$) and the middle matches the outer solution when $1-x \ll 1$ while $1-x \gg 1/N$.

\begin{figure}[h]
\begin{center}
\includegraphics[width=9cm]{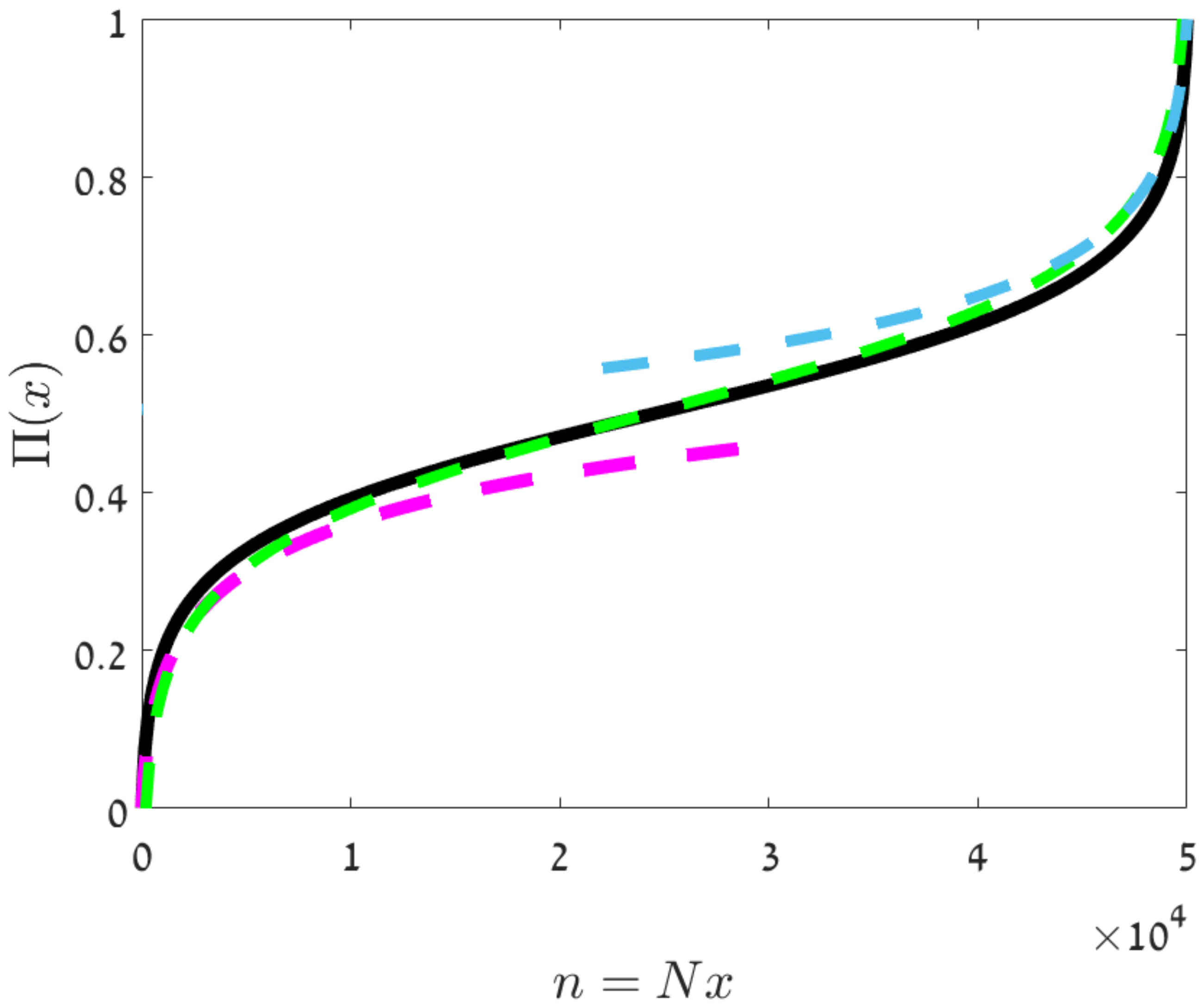}
\end{center}
\vspace{-0.5 cm}
\caption{The numerical solution of the BKE [Eq.~(\ref{exactnum}), with the transition probabilities in Eq.~(\ref{destinations})] (solid black line) is plotted against the solutions of Eq.~(\ref{eq30}) (dashed lines) as found in \cite{danino2018fixation,meyer2018noise} in the inner (pink, Eq.~(\ref{in}) below), middle [green, Eq.~(\ref{middle})] and outer [blue, Eq.~(\ref{out})] regimes. The parameters are $N = 50000$, $s_0 = 0$ and $\sigma = 0.1$, meaning that $s_0 \ll \sigma \ll 1$ which ensures that the diffusion approximation results work well. Note the regions of overlap between the inner-middle and the middle-outer solutions. \label{fig2}}
\end{figure}

The outcome of this procedure is demonstrated in Figure~\ref{fig2}, where the result of a numerical solution of the original, discrete BKE is compared with the expressions suggested in \cite{danino2018fixation,meyer2018noise} for a case where the diffusion approximation is valid.

\section{Combining asymptotic matching with the WKB approach: the small-$q$ sector} \label{apF}

Now we begin to derive the main analytical results of our work, by superimposing asymptotic matching on the  WKB approach.  We begin in the small-$q$ sector, by presenting our WKB technique in different $x$-regimes.

\textbf{In the middle regime}, demographic noise is negligible and $q$ satisfies
\begin{equation} \label{trans2}
e^{s_0 q_\text{mid}} \cosh(\sigma q_\text{mid}) = 1,
\end{equation}
so
\begin{equation} \label{mid3}
\Pi_\text{mid}(x) = C_3 + C_2 \left( \frac{x}{1-x} \right)^{q_\text{mid}}.
\end{equation}
If $q$ is  small, this implies
\begin{equation} \label{qmid}
q_\text{mid} = -\frac{2 s_0}{s_0^2 + \sigma^2}.
\end{equation}

\textbf{In the  inner regime}, $x \ll 1$, so $z \approx \ln x$ is negative and large. In this limit, $\cosh(z/2) \approx 1/(2\sqrt{x})$ and $\cosh(s + z) \approx \exp(-s)/(2x)$. Accordingly [see Eq.~(\ref{defB})],
\begin{equation}
B(s) \approx \frac{2x+e^{-s}}{\sqrt{Nx}} \approx  \frac{e^{-s}}{\sqrt{n}}.
\end{equation}
Plugging this into Eq.~(\ref{sige}), we get
\begin{equation}
\sigma_\text{e}^2 (n) = \sigma^2 + \frac{K(s_0)}{n},
\end{equation}
where
\begin{equation}\label{eq:K}
K(s_0) = e^{-2s_0}\cosh(2 \sigma).
\end{equation}
Therefore, in the inner regime we can use
\begin{equation} \label{qin_smallq}
q_\text{in} = -\frac{2 s_0 n}{(s_0^2 + \sigma^2)n + K(s_0)}.
\end{equation}

In the \textbf{outer regime} we will use the symmetry relation
\begin{equation} \label{outer1}
\Pi(x|_{s_0}) = 1-\Pi(1-x|_{-s_0}),
\end{equation}
which allows us to derive the outer and the inner solutions using the same formulas.

Armed with these expressions, we can now rederive the diffusion approximation, understand the assumptions on which it relies, and improve it in several ways.

\subsection{Rederivation  of past results obtained using the  diffusion approximation}

To rederive the results of~\cite{danino2018fixation,meyer2018noise}), we assume $s_0 \ll \sigma \ll 1$ so $K(s_0) \approx 1$. In the middle regime this implies
\begin{equation} \label{middle}
\Pi_\text{mid}(x) = C_3 + C_2 \left( \frac{x}{1-x} \right)^\frac{-2 s_0}{\sigma^2}.
\end{equation}

The inner solution becomes
\begin{equation} \label{appmyeq}
q_\text{in} = -\frac{2s_0}{\sigma^2 +1/n}.
\end{equation}
Accordingly,
\begin{equation} \label{Sm}
S_\text{in}(n) = \int q_\text{in}(z) dz  = \int q_\text{in}(n) \frac{dn}{n} = -\int dn \frac{2s_0}{\sigma^2 n +1} = -\frac{2s_0}{\sigma^2} \ln(\sigma^2 n +1).
\end{equation}

Adding $\exp(S_\text{in})$ to the constant solution and using the condition $\Pi(x=0)=0$, one finds in the inner regime (note $n=Nx$)
\begin{equation} \label{in}
\Pi_\text{in}(x) = C_1 \left(1-e^{S(n)}\right) = C_1 \left[1-(1+Nx\sigma^2)^{-2s_0/\sigma^2} \right].
\end{equation}
Finally, in the outer regime we utilize the symmetry of the problem, $\Pi(x|_{s_0}) = 1-\Pi(1-x|_{-s_0})$, to obtain
\begin{equation} \label{out}
\Pi_\text{out} = 1-C_4 \left\{1-\left[1+N(1-x)\sigma^2\right]^{2s_0/\sigma^2} \right\}.
\end{equation}

Eqs.~(\ref{middle}), (\ref{in}) and (\ref{out}) are the same equations obtained using the diffusion approximation in \cite{danino2018fixation,meyer2018noise}. The constants $C_1$ to $C_4$ are determined, as explained above, by matching (see the next subsection) and the expressions obtained fit the numerical results in the appropriate parameter regime well, as demonstrated in Figure~\ref{fig2}.

This derivation of the DA result assumes $s_0 \ll \sigma$.  Due to this assumption the results depend solely on $2s_\text{e}/\sigma_\text{e}^2$, which is the single parameter that is involved in a standard diffusion equation of the canonical form (\ref{canonical}),  $\mathbb{E}(\Delta z) \Pi' + \mathrm{Var}(\Delta z) \Pi''/2=0$.  Since $\sigma \le \sigma_\text{e}$, this assumption implies that the system is deep inside the small-$q$ sector for every $x$ (or $z$). However, the requirement for the small-$q$ sector, $s_\text{e}/\sigma_\text{e} < 0.25$, say, is much weaker than $s_0 \ll \sigma$, and in the next subsection we will show how to derive a better approximation in this sector.

\subsection{The small-$q$ approximation }

To obtain better expressions we relinquish the assumption $s_0 \ll \sigma \ll 1$, so that in the inner regime one needs to use Eq.~(\ref{qin_smallq}) for $q$ [instead of Eq.~(\ref{appmyeq})]. Integrating over $dz = dn/n$ we get
\begin{align}
S_\text{in}(x) = q_\text{mid} \ln (1+QNx),
\end{align}
where $q_\text{mid}$ was defined in (\ref{qmid}) and $Q \equiv (s_0^2 + \sigma^2)/K =(s_0^2 + \sigma^2)/[\exp(-2s_0)\cosh(2\sigma)]$. The inner solution thus takes the form
\begin{equation} \label{pin3}
\Pi_\text{in}(x) = C_1\left[1-(1+QNx)^{q_\text{mid}}\right].
\end{equation}
The symmetry relation (\ref{outer1}) implies
\begin{equation}\label{pout3}
\Pi_\text{out}(x) = 1-C_4 \left\{1-\left[1+\tilde{Q}N(1-x)\right]^{-q_\text{mid}}\right\}.
\end{equation}
where $\tilde{Q} \equiv Q(-s_0) = (s_0^2+\sigma^2)/[\exp(2s_0)\cosh(2\sigma)]$. The middle solution is, like Eq.~(\ref{middle}) but without the $s0 \ll \sigma$ restriction,
\begin{equation} \label{pmid3}
\Pi_\text{mid}(x) = C_3 + C_2 \left( \frac{x}{1-x} \right)^{q_\text{mid}}.
\end{equation}

Matching the inner and the middle solutions in the region $1/N \ll x \ll 1$, and the middle and the outer solutions in the region $1/N \ll 1-x \ll 1$, we get
\begin{equation}
C_1-C_1(QNx)^{q_\text{mid}} = C_3+C_2 x^{q_\text{mid}} \qquad \text{and} \qquad C_3+C_2(1- x)^{-q_\text{mid}} = 1-C_4+C_4\left[\tilde{Q}N(1-x)\right]^{-q_\text{mid}}.
\end{equation}
This gives the required constants,
\begin{equation} \label{Cs}
C_1 = C_3 = \frac{1}{1-\left(\text{N}^2 Q \tilde{Q} \right)^{q_\text{mid}}}, \qquad C_2  = \frac{(\text{N} Q)^{q_\text{mid}}}{\left(\text{N}^2 Q \tilde{Q}\right)^{q_\text{mid}}-1} \qquad \text{and} \qquad C_4 =  \frac{\left(\text{N}^2 Q \tilde{Q}\right)^{q_\text{mid}}}{\left(\text{N}^2 Q \tilde{Q}\right)^{q_\text{mid}}-1}.
\end{equation}

\begin{figure}[h]
\centering
\includegraphics[width=0.85\textwidth]{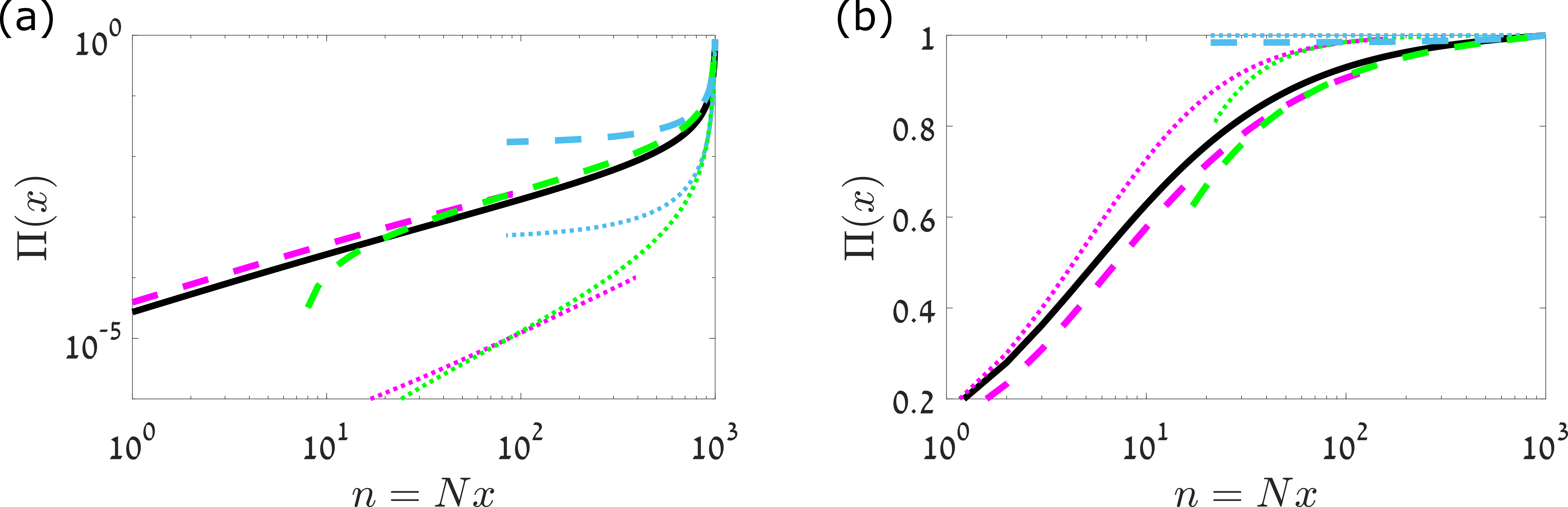}
\caption{\label{fig:improved_DA} Chance of fixation $\Pi(x)$ plotted vs.\ $x$. The outcome of a direct numerical solution of the BKE is depicted as a solid black line. The dashed lines correspond to the inner (pink), middle (green) and outer (blue) solutions of the small $q$ approximation, i.e., to Eqs.~(\ref{pin3}), (\ref{pmid3}) and (\ref{pout3}) with the constants $C_1$ to $C_4$ as obtained in Eq.~(\ref{Cs}). The dotted lines, with the same color code, represent the diffusion approximation expressions (\ref{in}), (\ref{middle}) and (\ref{out}). In the left panel $s_0$ is negative and the chance of fixation is small, so a double-logarithmic scale is used. In the right panel $s_0$ is positive and the results are plotted using a semi-logarithmic scale. In both cases the small-$q$ approximation (dashed) results fit quite well the direct numerical (black) curve: the inner solution works well for small $x$, the outer one when $x$ is close to $1$ and the middle one in the middle regime, with noticeable overlap between these curves in the domains where the matching takes place. On the other hand, the diffusion approximation (dotted) curves do not fit the correct result and (in some cases, like the inner-middle regime of the left panel) fail to match each other. The parameter values are $\sigma = 0.5$; $N = 1000$; and $s_0 = -0.1$ (left) and $s_0 = 0.1$ (right).}
\end{figure}

Figure~\ref{fig:improved_DA} compares the results of the direct numerical solution of the BKE with the analytical expressions of the diffusion approximation and  the improved small-$q$ approximation, in all the three regimes, for beneficial (positive $s_0$) and deleterious (negative $s_0$) mutants. Clearly, the improved small-$q$ approximation does a much better job of reproducing the numerical result.

\begin{figure}[h]
\centering
\includegraphics[width=0.98\textwidth]{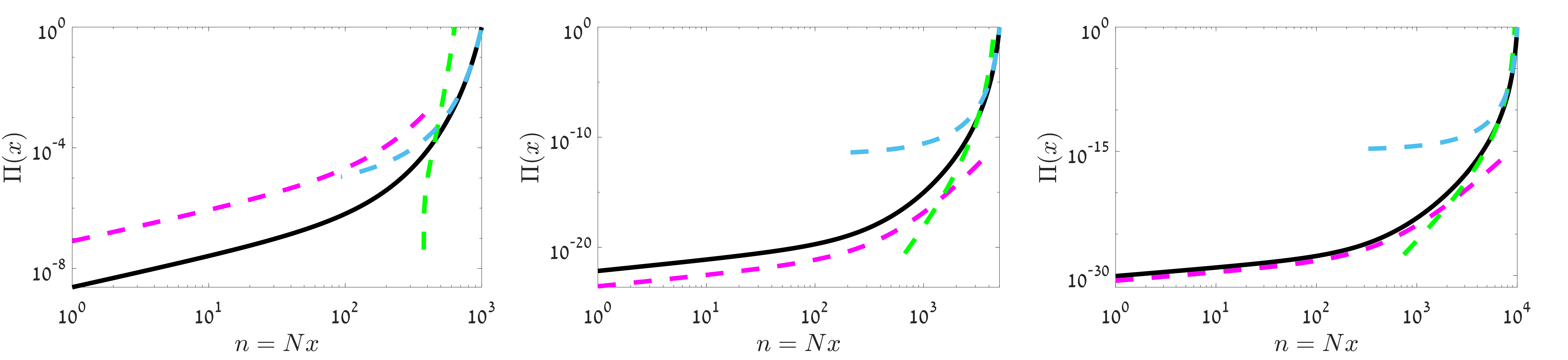}
\caption{\label{imp} The performance of the analytical expressions obtained using the small-$q$ approximation improves as $N$ increases. Here the direct numerical results (solid black) are compared with the inner (dashed pink), middle (dashed green), and outer (dashed blue) solutions of the improved small-$q$ approximation for different values of $N$ (left panel: $N = 1000$,  middle panel: $N = 5000$, right panel: $N = 10000$). While for $N=1000$ the improved small-$q$ curves perform poorly, they fit the numerical curve very well when $N =10000$. The other parameters are $s_0 = -0.01$ and $\sigma = 0.04$.}
\end{figure}

The asymptotic matching approach relies on $N$ being large. The larger the value of $N$, the wider is the middle regime where demographic stochasticity is negligible, and wider correspondingly is the overlap between this regime and the inner/outer regimes. One therefore expects the analytical expressions to fit the numerical results better as $N$ increases, and this is indeed what happens, as demonstrated in Figure~\ref{imp}.

Unlike the DA, which depends on the ratio between the mean and the variance of $\Delta z$, in the small-$q$ approximation the important parameter is the ratio between the mean and the second moment. As explained in Section~\ref{DAs}, the procedure that yields the diffusion approximation does not allow one to distinguish between the variance and the second moment, since the justification of the continuum approximation requires the mean to be negligible with respect to the second moment. Here, the use of the WKB approach makes it clear that it is the second moment, and not the variance, that is the correct coefficient of the $\Pi''$ term in the small-$q$ sector.

\subsection{The chance of fixation of a single mutant}

Our WKB approach is based on the assumption that the logarithm of $\Pi$ is smooth over the integers. When the number of individuals is very small (one or two, say), even this assumption may break down. It is nevertheless instructive to examine the predictions of our formula for the case of a single mutant. Even when the results in this case are not fully accurate, they still provide a correct order of magnitude estimation with small relative errors.

\begin{figure}[h]
\centering
\includegraphics[width=0.45\textwidth]{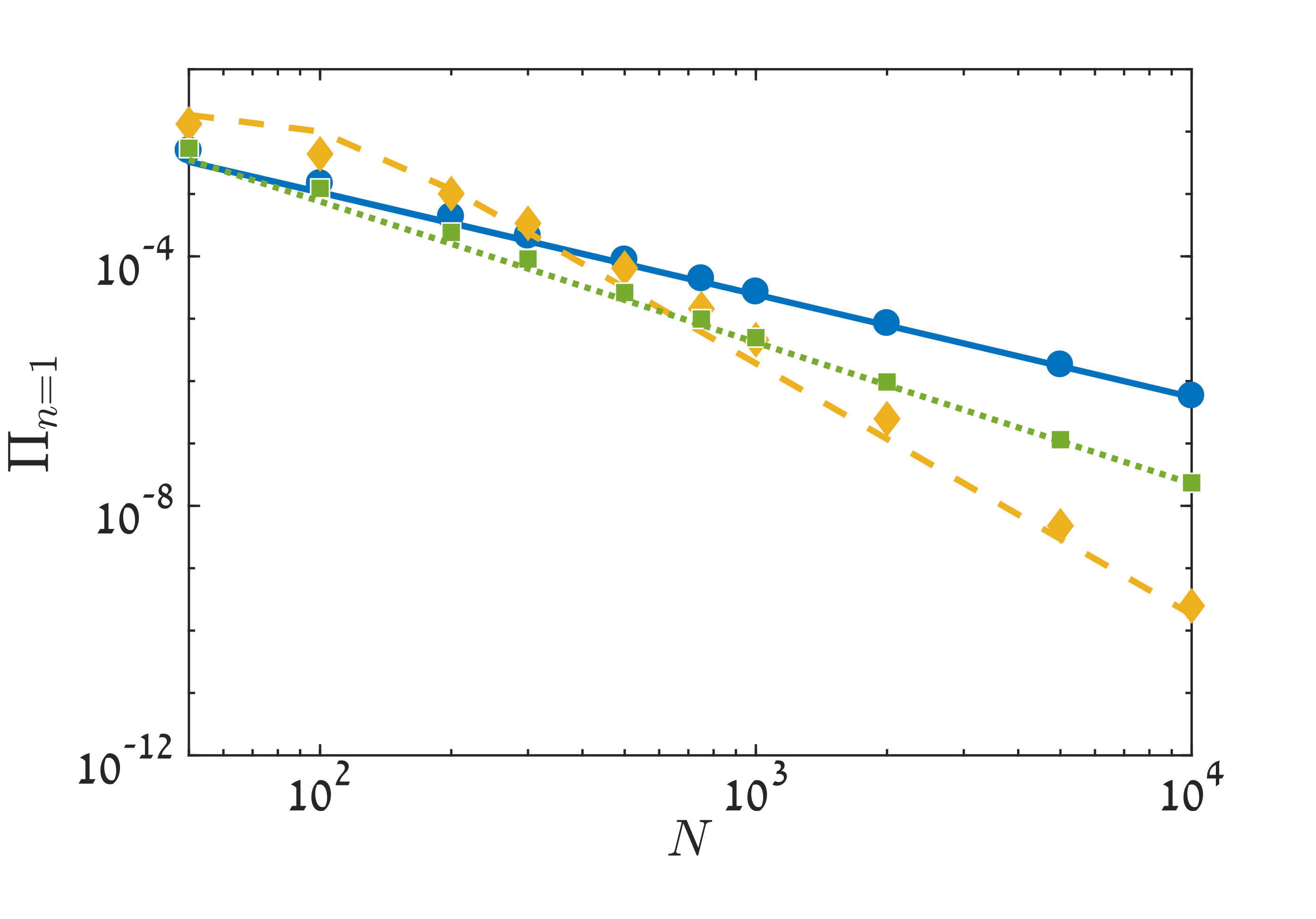}
\caption{\label{single8} The chance of fixation of a single mutant, $\Pi_{n=1}$, is plotted vs.\ $N$. Symbols are the results of direct numerical solution, while lines are the prediction of Eq.~(\ref{single4}). The parameters are $s_0 = -0.1$, $\sigma = 0.5$ (full line, circles); $s_0 = -0.05$, $\sigma = 0.3$ (dotted line, triangles); and $s_0 = -0.01$, $\sigma = 0.1$ (dashed line, diamonds). The theory captures the general trend well, and the relative error decreases with $N$. To improve the quality of the approximation, the parameter $q_\text{mid}$ has been taken from the direct numerical solution of Eq.~(\ref{mtrans}) with $s_\text{e} = s_0$ and $\sigma_\text{e} = \sigma$.}
\end{figure}

Eqs.~(\ref{pin3}) and (\ref{Cs}) yield the chance of a single mutant to reach fixation,
\begin{equation} \label{single4}
\Pi_{n=1} =  \frac{1-\left[1+\frac{e^{2s_0}(s_0^2+\sigma^2)}{\cosh(2\sigma)}\right]^{-\frac{2s_0}{s_0^2+\sigma^2}}}
{1-\left[\frac{N(s_0^2+\sigma^2)}{\cosh(2\sigma)}\right]^{-\frac{4s_0}{s_0^2+\sigma^2}}}.
\end{equation}
Figure~\ref{single8} shows that this formula works quite well.

It is interesting to examine the $N$-dependence of $\Pi_{n=1}$. When $q_\text{mid}$ is negative (i.e.\ for a beneficial mutant, with $s_0>0$), the $N$-dependent term in the denominator disappears when $N \to \infty$, and the chance of fixation becomes  $N$-independent. This implies that above some critical value $n_\text{c}$ (see next subsection) the chance of fixation is nearly one. On the other hand, when $q_\text{mid}$ is positive (i.e.\ for a deleterious mutant, with $s_0<0$), the chance of fixation decays to zero when $N \to \infty$.

When selection is fixed [Eq.~(\ref{sol})] this decay is exponential, but in our case the decay of  $\Pi$ is only as a power law in $N$, $\Pi \sim N^{-2q_\text{mid}}$. This happens because selection may change  sign: although $s_0<0$, as long as $\sigma > |s_0|$ there are periods when $s = s_0+\sigma>0$. Accordingly, the mutant may reach fixation due to an improbable series of good years. The length of such a series that leads to fixation is logarithmic in $N$, so its probability decays like a power law~\cite{yahalom2019comprehensive}. On the other hand, when the system does not allow good years the mutant can win only due to demographic noise (as it needs a rare series of binomial trials in which its actual abundance grows despite the expected abundance decreasing), and then the chance of fixation decays exponentially with $N$.

\subsection{Weak and strong selection in fluctuating environments}

An important aspect of our analysis has to do with the distinction between weak and strong selection. In a fixed environment, if $|s|N \ll 1$ (weak selection) the dynamics are neutral. If $|s|N \gg 1$ (strong selection) the dynamics are still approximately neutral (dominated by demographic stochasticity) up to $n_\text{c} = 1/(2|s|)$ and are nearly deterministic above this point, so the chance of fixation for a beneficial mutant becomes $N$-independent for $N>n_\text{c}$~\cite{desai2007speed}. For a deleterious mutant, $\Pi$ is exponentially small in $N$ as long as $N>n_\text{c}$.

Analogously, in a stochastic environment $n_\text{c}$ is the point at which  $S_n  \approx 1$ where  $S = \int q dz = \int (q/n) dn  $. We assume that $n_\text{c}$ happens to lie in the inner regime where $z = \ln x$ (but still $n_\text{c} \gg 1$).  In the inner regime $q$ is usually small, so $S \approx 2s_0 \ln[1+n(s_0^2+\sigma^2)]/(s_0^2 + \sigma^2)$. As a result,
\begin{equation}
n_\text{c} = \frac{e^\frac{s_0^2+\sigma^2}{2s_0}-1}{s_0^2+\sigma^2}.
\end{equation}
This expression generalizes a similar criterion suggested in~\cite{cvijovic2015fate}. The two expressions coincide when the DA holds, i.e., when $s_0^2$ is neglected with respect to $\sigma^2$.  The fixed-environment expression $1/(2|s_0|)$ emerges when $\sigma \to 0$. $n_\text{c}$ diverges exponentially when $|s_0| \to 0$, meaning that even huge populations may be in the weak selection regime. In general, to obtain an expression for $n_\text{c}$ when our assumptions do not hold, one has to calculate $S$ for a given $q$ and to apply the condition $S(n_\text{c}) =1$.

Importantly, in a stochastic environment weak selection does not imply neutrality. Instead, in the weak-selection regime the system is in a ``time-averaged neutral"~\cite{kalyuzhny2015neutral,danino2018theory} phase, meaning that $s_\text{e}$ is effectively zero so both types have the same mean fitness (when  fitness is averaged over time). In this case the abundance undergoes an unbiased random walk along the $z$-axis, so the chance of fixation at $n$ behaves in general like $z/z_\text{max}$ and decays logarithmically with $N$. Only when $\sigma \sqrt{N} \ll 1$ does the demographic noise dominate the fluctuating selection and Kimura's neutral dynamics (in which the chance of fixation is $n/N$) are restored.

\section{Beyond the small-$q$ sector } \label{lastapp}

In the last section we obtained analytical expressions for $\Pi$ in the small-$q$ sector, i.e, when $|\tilde{s}| = |s_\text{e}/\sigma_\text{e}|<0.25$ for all values of $z$ (or $x$). In this section we aim to expand our analytical technique to the case where this condition is not satisfied. We assume that $N$ is large enough such that the middle regime, in which demographic stochasticity is negligible, exists (see Section~\ref{middle5} above). In general, since $\sigma_\text{e}$ increases towards the extinction/fixation points, we expect that $|\tilde{s}| > 0.25$ in this middle regime, while in the inner and the outer regimes the small-$q$ approximation still holds. Accordingly, in the middle regime we still have
\begin{equation}
\Pi_\text{mid} (x) = C_3 + C_2 \left(\frac{x}{1-x}\right)^{q_\text{mid}},
\end{equation}
but now $q_\text{mid}$ is obtained (using Table \ref{table1}) from the fundamental transcendental equation $\exp(q s_0)\cosh(q \sigma)=1$, and we do not assume [as in Eq.~(\ref{qmid})] that $q$ is small.

In the inner and outer regimes $\sigma_\text{e}$ is larger, so one may try to employ the small-$q$ solutions obtained in the last section, with minimal modification to ensure that their large-$n$ limit matches the solution in the middle regime. To do this we reconsider the expression for $q_\text{in}$ in the small-$q$ regime in Eq.~(\ref{qin_smallq}). That can be rewritten as
\begin{align}
q_\text{in} = -\frac{2 s_0 n}{-\dfrac{2s_0 n}{q_\text{mid}} + K(s_0)},
\end{align}
with $K(s_0)$ defined in Eq.~(\ref{eq:K}) and where $q_\text{mid} = -2s_0/(s_0^2 + \sigma^2)$ is the value of $q$ in the middle regime when $q$ is small. Analogously, for the case when $q$ is no longer always small, we suggest the approximation
\begin{equation}\label{eq:q0}
q_\text{in} = \dfrac{-2s_0 n}{K(s_0)-\dfrac{2s_0 n}{q_\text{mid}}},
\end{equation}
with the same expression for $K(s_0)$ as before, and with $q_\text{mid}$ now being the  solution of the fundamental transcendental equation $\exp(q s_0)\cosh(q \sigma)=1$. Such an approximation for $q_\text{in}$ in the present case (when $q$ is not always small) helps in the sense that Eq.~(\ref{eq:q0}) converges to $q_\text{mid}$ when $n$ is large and to $-2s_0n/K(s_0)$ when $n$ is small.

The rest of the calculation proceeds along the lines of the last section, and the results turn out to be exactly those obtained in Eq.~(\ref{Cs}), with the substitutions
\begin{equation}
Q \to -\frac{2s_0}{K(s_0) q_\text{mid}} \qquad \text{and} \qquad \tilde{Q} \to -\frac{2s_0}{K(-s_0) q_\text{mid}}.
\end{equation}

\begin{figure}[h]
\centering
\includegraphics[width=0.8\textwidth]{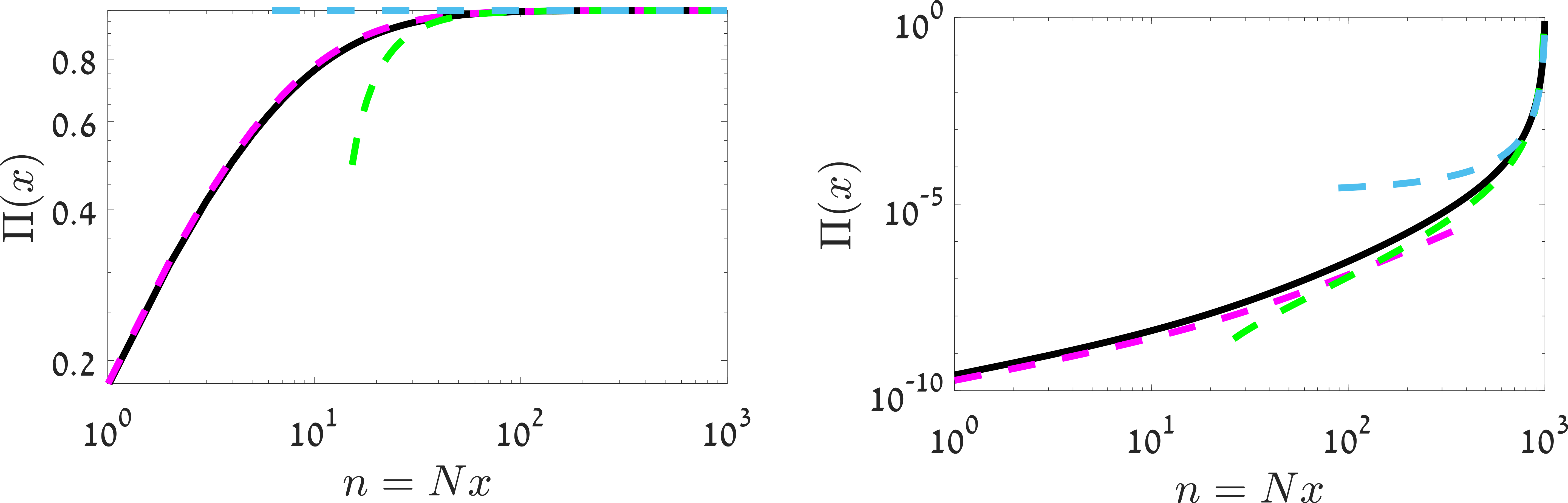}
\caption{\label{fig:with_q0} Chance of fixation as found from the direct numerical solution of the BKE (solid black curve), and using our analytical approximation for the regimes where $q$ is not small (dashed curves), for a beneficial (left panel) and a deleterious (right panel) mutant. The inner (pink), middle (green) and outer (blue) solutions cover the numerical answer well and show good overlap. The parameters are $\sigma = 0.3$; $N = 1000$; and $s_0 = 0.1$ (left panel) and $s_0 = -0.1$ (right panel). In both cases $|\tilde{s}| \approx 0.35$, so the small-$q$ approximation does not work.}
\end{figure}

Figure~\ref{fig:with_q0} shows a comparison between the numerical solution of the BKE (solid black curve), and the semi-analytical curves in the different regimes.

\section{Discussion}

The analysis of  competition between different types (strains, alleles, species) lies at the heart of the theory of population genetics, ecology and evolution. The quantity we have considered here, the chance of ultimate fixation, together with other quantities like  the mean time to absorption (either fixation or loss), determines the genetic polymorphism in a population and the divergence rate between populations~\cite{ewens2012mathematical}, marks the transition between the successive fixation and clonal interference phases of evolutionary dynamics~\cite{desai2007speed}, and controls the species richness in ecological communities~\cite{azaele2016statistical}. Given the increasing recognition of the importance of environmentally-induced selection fluctuations in the wild~\cite{bergland2014genomic,bell2010fluctuating,messer2016can,caceres1997temporal,hoekstra2001strength,leigh2007neutral,hekstra2012contingency,kalyuzhny2014niche,kalyuzhny2014temporal,chisholm2014temporal}, many studies have focused on the calculation of these quantities in a varying environment.

The standard tool in the analysis of such a problem is the diffusion approximation. Technically (see Appendix \ref{app:ignoringPi3}), this approximation involves an expansion of $\Pi(x)$ in a Taylor series in $1/N$, with the series truncated after the second order. The neglect of the higher-order terms in the series reflects the assumption that $\Pi$ is smooth over the integers. As a result of this truncation, the answer depends only on the first two moments of the walk. In the WKB approximation suggested here we discard higher derivatives of $\ln \Pi$, not of $\Pi$, so the smoothness requirements are much weaker. Our two-destination procedure replaces the actual walk by an effective, simpler walk that preserves the first two moments of the original process. Accordingly, while our method takes into account higher-order terms in the expansion of $\Pi$ (since it is the \emph{logarithm} of $\Pi$ that is truncated at the second order), it still relies on only the first two moments.

To make the problem analytically tractable, it is helpful to integrate out (average over) the environmental degrees of freedom and to write an effective one-dimensional equation for the population process. This is a justifiable approach when the temporal scale associated with environmental variations is much shorter than the fixation time. Within the DA framework, this was the strategy used in old and recent studies~\cite{takahata1975effect,takahata1979genetic,danino2018fixation,meyer2018noise}. In our method, the averaging over the fast degrees of freedom is hidden in the two-destination approximation.

The diffusion approximation, which shows a remarkably good performance when applied to systems under fixed selection, has a rather narrow range of applicability for systems with fluctuating selection. We have shown here that the controlling-factor WKB method has a much wider range of applicability.

Our method involves three approximation steps. First, we employ the \emph{WKB approximation} that depends on much weaker smoothness requirements. Second, we use the \emph{two-destination approximation} which provides a one-dimensional system and a tractable transcendental equation. Finally, when $N$ is large enough and the middle regime exists, we employ the \emph{asymptotic matching technique}, solved in the inner, outer and middle regimes, and match the outcomes in the regions of overlap.

Each of these three steps must be examined in itself. The third step, asymptotic matching, is a well-established and trusted technique, and it is known that the corrections scale as powers of the small parameter $1/N$. The controlling-factor WKB method neglects physical optics corrections; our preliminary calculations (not included in this paper) suggest that the related corrections disappear like inverse powers of $N$ as well. The two-destination approximation is more subtle, as it neglects long jumps and puts a limit on the applicability of the technique presented here. The replacement of all possible jumps by two destinations, $s_\text{e} + \sigma_\text{e}$ and $s_\text{e} - \sigma_\text{e}$, implies that once $\sigma_\text{e}$ becomes smaller than $|s_\text{e}|$, the jumps become unidirectional and the system flows deterministically to either fixation or extinction, whereas in practice rare long jumps may still lead to a different result. Because of this fact our approximation breaks down as $\sigma \to |s_0|$, where the relative importance of these long jumps increases. To overcome this difficulty, one may either abandon the two-destination approximation (in which case a more complicated transcendental equation appears), or devise an alternative two-destination scheme in which long jumps are taken into account while keeping the same mean and variance as $s_\text{e}$ and $\sigma_\text{e}^2$, respectively. We intend to employ such a strategy in a forthcoming work to address analytically the cases where $\sigma$ is close to, or even smaller than, $|s_0|$.

In the small-$q$ sector, when for any $z$ the ratio $|\tilde{s}| = |s_\text{e}/\sigma_\text{e}|$ is smaller than $0.25$, we provide in Section~\ref{apF} a complete asymptotic matching solution. The main novelty in this solution, with respect to the standard diffusion approximation, is the replacement of the parameter $s_0/\mathrm{Var}[\Delta z]$ by the ratio of $s_0$ and the second moment of $\Delta z$, $s_0/\mathbb{E}[(\Delta z)^2]$. Note that the use of the variance is not a necessary part of the DA formalism; as explained in Section \ref{fixed}, when the assumptions behind the DA hold, the difference between the second moment and the variance is negligible, so each of them may serve equally well as the coefficient of the stochastic term $\Pi''$. The variance is frequently used in the literature~\cite{karlin1981second}, since intuitively one expects the strength of the stochasticity to be proportional to the variance -- a parameter that reflects how much the jumps are scattered around their mean -- and not to the second moment which partially includes the deterministic bias. However, as we have seen here, in our case the WKB analysis  promotes the use of the second moment, and indeed the expressions obtained for $\Pi$ in the small-$q$ approximation fit the results much better than the variance-based DA expression.

When $q$ in the middle regime is not small, as in the case discussed in Section~\ref{lastapp}, the application of the asymptotic matching approach becomes more complicated. Deep in the inner and outer regimes the strength of demographic stochsticity is strong and the solution belongs to the small-$q$ sector, whereas in the middle regime it must approach other values. Here we presented a simple \emph{ad hoc} solution to this problem by devising an expression that converges to the appropriate limit on both sides, neglecting possible mismatches between these two ends. Better approximations may be based on more detailed matching techniques in which the number of different regions increases.

In this work we have considered only dynamics that have no attractive fixed points except the absorbing states at extinction and fixation. When the system admits an attractive fixed point at finite $x$, either because of density-dependent feedback or due to stochasticity-induced stable coexistence~\cite{chesson1981environmental,chesson2000mechanisms,dean2020stochasticity}, the chance of invasion or establishment becomes a very important quantity. The properties of the chance of invasion in this case may differ significantly from those of the chance of fixation; for example, the chance of invasion may be quite large despite the chance of fixation being very small, meaning that the system supports the long-lasting transient existence of ultimately-extinct species~\cite{meyer2018noise}. Nevertheless, as a technical problem, the calculation of this quantity, the chance of invasion, is very similar to the calculation of $\Pi$ as detailed here. We intend to exploit this feature in future work.

\noindent \textbf{Acknowledgments:}
This research was supported by the ISF-NRF Singapore joint research program (grant number 2669/17).
\clearpage
\appendix

\section{The assumptions behind the diffusion approximation for populations under fixed selection}\label{app:ignoringPi3}

The derivation of the diffusion approximation for random walks and other stochastic processes, either through the Kramers-Moyal expansion or via van Kampen's $\Omega$-expansion, is an established procedure~\cite{gardiner1985handbook}. In section~\ref{fixed} we appplied a Kramers-Moyal-type expansion to Eq.~(\ref{BKE}); here we would like to clarify the conditions for its validity by monitoring carefully the neglected higher-order terms.

We consider Wright-Fisher (non-overlapping-generation) dynamics under fixed selection. As stated in Eq.~(\ref{smooth}), the assumed smoothness of $\Pi_n$ over the integers $n$ allows one to approximate $\Pi_{n+m} = \Pi(x+m/N)$, where $x = n/N$, as
\begin{equation} \label{smooth_from_main}
\Pi(x+m/N) = \Pi(x) + \frac{m}{N} \Pi'(x) + \frac{m^2}{2N^2} \Pi''(x) + \frac{m^3}{6N^3} \Pi'''(x) + \rm{higher \ order \  terms},
\end{equation}
where primes denote derivatives with respect to $x$. As stated in Eq.~(\ref{destinations1}), the transition probability $W_{n \to n+m}$ to go from a population of size $n$ to a population of size $n+m$ (where $n+m$ may range from $0$ to $N$) in a single step is
\begin{equation} \label{destinations1_from_main}
W_{n \to n+m} = \binom{N}{n+m} r^{n+m}(1-r)^{N-n-m},
\end{equation}
where [see Eq.~(\ref{eq01})] $r$ is the probability of the mutant type to capture any given slot out of the $N$ open slots in each generation, namely
\begin{equation} \label{eq01_from_main}
r = \frac{x e^s}{x e^s + (1-x)}.
\end{equation}

The first three moments of the change in population ($m$) are therefore:
\begin{align}\label{eq:m_moments}
\overline{m} &= \sum\limits_{m=-n}^{N-n} m W_{n \to n+m} = N(r-x) \nonumber\\
	     &= Nsx(1-x) + \frac{Ns^2x(1-x)}{2} + {\cal O}\left(s^3\right),\nonumber\\
\overline{m^2} &= \sum\limits_{m=-n}^{N-n} m^2 W_{n \to n+m} = N^2(r-x)^2 + Nr(1-r) \nonumber\\
	       &= Nx(1-x) + Nsx(1-x)(1-2x) + \frac{Ns^2x(1-x)\left[1+2x(1-x)(N-3)\right]}{2} + {\cal O}\left(s^3\right), \text{ and}\\
\overline{m^3} &= \sum\limits_{m=-n}^{N-n} m^3 W_{n \to n+m} = N^3(r-x)^3 + 3rN^2(r-x)(1-r) + Nr(1-r)(1-2r) \nonumber\\
	       &= Nx(1-x)(1-2x) + Nsx(1-x)\left[1+3x(1-x)(N-2)\right] + \frac{Ns^2x(1-x)(1-2x)(1+3x(1-x)(3N-4))}{2} + {\cal O}\left(s^3\right).\nonumber
\end{align}

Here note that, for any $i$, the highest power of $N$ in the ${\cal O}\left(s^3\right)$ terms in $\overline{m^i}$ is $i$, so if we wish to look at only the highest-magnitude terms, we can forget all the other terms hidden in ${\cal O}\left(s^3\right)$ and consider only the $N^i s^3$ terms. (This statement assumes that $N$ is large and $s$ is small, in some way; these notions will be made precise below.)

Putting in Eqs.~(\ref{smooth_from_main}) and (\ref{eq:m_moments}) in the Backward Kolmogorov Equation [Eq.~(\ref{BKE})], we get
\begin{align}\label{eq:DAprecursor}
0 = & \frac{\overline{m}}{N}\Pi' + \frac{\overline{m^2}}{2N^2}\Pi'' + \frac{\overline{m^3}}{6N^3}\Pi''' + ...\nonumber\\
  = & \left\{sx(1-x) + \frac{s^2x(1-x)}{2} + A(x) s^3\right\}\Pi' \\
    & + \left\{\frac{x(1-x)}{2N} + \frac{sx(1-x)(1-2x)}{2N} + \frac{s^2x(1-x)\left[1+2x(1-x)(N-3)\right]}{4N} + B(x) s^3\right\}\Pi'' \nonumber\\
    & + \left\{\frac{x(1-x)(1-2x)}{6N^2} + \frac{sx(1-x)\left[1+3x(1-x)(N-2)\right]}{6N^2} + \frac{s^2x(1-x)(1-2x)(1+3x(1-x)(3N-4))}{12N^2} + C(x) s^3\right\}\Pi'''\nonumber\\
    & + ...\nonumber,
\end{align}
where $A(x)$, $B(x)$, $C(x)$, etc., are functions only of $x$, not of $N$. This is because, as explained above, the largest coefficient of $s^3$ (and $s^4$, $s^5$, etc.) in $\overline{m^i}$ is proportional to $N^i$, so division of $\overline{m^i}$ by $N^i$ makes the $N$-dependence in these terms disappear. We have ignored the other, smaller-magnitude, terms.

$N$ is by assumption a large number, $N \gg 1$. In order to reduce Eq.~(\ref{eq:DAprecursor}) to the diffusion equation, which in the current context takes the form~\cite{kimura1962probability,crow1970introduction,ewens2012mathematical}
\begin{align}\label{eq:DAsimple}
\Pi'' + 2Ns\Pi' = 0,
\end{align}
the only other assumption needed is $s$ being small enough such that
\begin{align}
Ns^2 \ll 1.
\end{align}
With this assumption, checking Eq.~(\ref{eq:DAprecursor}) term-by-term, it becomes clear that the dominant terms are
\begin{align}\label{eq:DAreduction}
0 = sx(1-x)\Pi' + \frac{x(1-x)}{2N}\Pi'',
\end{align}
which is the same as Eq.~(\ref{eq:DAsimple}).

Under the above assumption, the second moment of $m$ reduces to its variance. This is seen on calculating the variance explicitly,
\begin{align}\label{eq:Var_m}
\mathrm{Var}(m) &= \overline{m^2} - \overline{m}^2 = N r (1-r) \nonumber\\
	        &= Nx(1-x) + Nsx(1-x)(1-2x) + \frac{Ns^2x(1-x)\left[1-6x(1-x)\right]}{2} + {\cal O}\left(s^3\right).
\end{align}
Comparison of this with Eq.~(\ref{eq:m_moments}) shows that $\mathrm{Var}(m)$ and $\overline{m^2}$ differ only in the $s^2$ terms (and those with higher powers of $s$), which are smaller than $\overline{m^2}$ precisely by a factor of $Ns^2 \ll 1$.

\bibliography{refs_ives}

\end{document}